\documentclass{iopart}
\usepackage{graphicx}  
\usepackage{latexsym}  

\jl{6}        
\eqnobysec    

\def\beq{\begin{equation}}
\def\eeq{\end{equation}}

\def\rmd{{\rm d}}
\def\sgn{\mathop{\rm sgn}}

\def\diag{\mathop{\rm diag}}
\def\II{I} \def\JJ{J}  

\def\rmd{{\rm d}}
 
\def\dual{{}^{*}\kern-1pt }

\def\selfd{\,\mathop{\tilde{\,}}\nolimits\,}
\def\leftselfd{\selfd\kern-1.7pt}

\begin{document}

\title[EKG of the Mixmaster Universe]
{Electrocardiogram of the Mixmaster Universe}

\author{
Donato Bini$^* {}^\S{}^\P$, Christian Cherubini$^\dagger$,
Andrea Geralico$^\ddag {}^\S$ and Robert T. Jantzen$^{**}$}
\address{
  ${}^*$\
Istituto per le Applicazioni del Calcolo ``M. Picone,'' CNR I-00161 Rome, Italy
}
\address{
  ${}^\S$\
  ICRA, University of Rome ``La Sapienza,'' I-00185 Rome, Italy and
  ICRANET, I--65122 Pescara, Italy
}

\address{
${}^\P$
  INFN - Sezione di Firenze, Polo Scientifico, Via Sansone 1, 
  I-50019, Sesto Fiorentino (FI), Italy 
}

\address{
$^\dagger$
Nonlinear Physics and Mathematical Modeling Lab, University Campus Bio-Medico,
I--00128 Rome, Italy
}

\address{
  ${}^\ddag$\
  Physics Department, University of Rome ``La Sapienza,'' I-00185 Rome, Italy
}

\address{
  ${}^{**}$\
Department of Mathematical Sciences, Villanova University, Villanova, PA 19085  USA
}

\begin{abstract}
The Mixmaster dynamics is revisited in a new light as revealing a series of transitions in the complex scale invariant scalar invariant of the Weyl curvature tensor best represented by the speciality index $\mathcal{S}$, which gives a 4-dimensional measure of the evolution of the spacetime independent of all the 3-dimensional gauge-dependent variables except for the time used to parametrize it. Its graph versus time characterized by correlated isolated pulses
in its real and imaginary parts corresponding to curvature wall collisions serves as a sort of electrocardiogram of the Mixmaster universe, with each such 
pulse pair arising from a single circuit or ``complex pulse'' around the origin in the complex plane.
These pulses in the speciality index and their limiting points on the real axis seem to invariantly characterize some of the so called spike solutions in inhomogeneous cosmology and should play an important role as a gauge invariant lens through which to view current investigations of inhomogeneous Mixmaster dynamics.
\end{abstract}

\pacno{04.20.Cv}

\section{Introduction}

The Bianchi type IX spatially homogeneous vacuum spacetime also known as the Mixmaster universe \cite{misner69}
has served as a theoretical playground for many ideas in general relativity, one of which is the question of the nature of the chaotic behavior exhibited in some solutions of the vacuum Einstein equations \cite{hobillbook} and another is the question of whether or not one can interpret the spacetime as a closed gravitational wave \cite{wheeler,king}. In particular, to describe the mathematical approach to an initial cosmological singularity, the exact Bianchi type IX dynamics leads to the BKL (Belinski-Khalatnikov-Lifshitz \cite{BKL}) approximation involving the discrete BKL map which acts as the transition between phases of approximately Bianchi type I evolution. The parameters of this map are not so easily extracted from the numerical evolution of the metric variables \cite{berger1994,berger1997}. However, recently it has been realized that these parameters are directly related to transitions in the scale invariant part of the Weyl tensor \cite{bcj2007a,bcj2007b}. In fact this leads to a whole new interpretation of what the 
discrete
BKL dynamics represents, by lifting its interpretation from transitions in apparently gauge-dependent slicing geometry quantities to transitions in gauge-independent spacetime curvature invariants.

For a given 
spacelike slicing 
of any spacetime, one can always introduce the scale invariant part of the extrinsic curvature when its trace is nonzero by dividing by that trace, thus reducing the information it carries to be equivalent to that contained in the scale invariant part of the shear tensor, which describes the anisotropy in the cosmological evolution of the spatial geometry of the time coordinate surfaces. In the expansion-normalized approach to spatially homogeneous 
dynamics, this corresponds to the expansion-normalized gravitational velocity variables \cite{ds}. This scale invariant extrinsic curvature tensor can be characterized by its eigenvalues, whose sum is 1 by definition: these define three functions of the time parametrizing the foliation which generalize the Kasner indices of Bianchi type I vacuum spacetimes 
where they reduce to constants. 
We will refer to them as the generalized Kasner indices following Belinski and Francaviglia, who refer to the corresponding orthogonal spatial frame of eigenvectors as the generalized Kasner axes \cite{belinski}. 
These quantities are defined for any timelike slicing of a completely general spacetime, where they will be functions of both the time and spatial coordinates.
A phase of ``velocity-term-dominated" evolution \cite{eardley,isemon1990}  (also referred to as Kasner evolution) is loosely defined as an interval of time during which the spatial curvature terms in the spacetime curvature are negligible compared to the extrinsic curvature terms. Under these conditions the vacuum Einstein equations can be approximated by ordinary differential equations in the time. These lead to a simple scaling of the eigenvectors of the extrinsic curvature during which the generalized Kasner indices remain approximately constant and simulate the Bianchi type I spatially flat spatially homogeneous Kasner evolution.

The Weyl tensor can be also be repackaged as a second rank but complex spatial tensor with respect to the foliation \cite{ES} and its scale invariant part is determined by a single complex scalar function of its eigenvalues, a number of particular representations for which are useful. In particular the so called speciality index \cite{bakcam2000} is the natural choice for this variable which is independent of the permutations of the spatial axes used to order the eigenvalues, and so is  a natural 4-dimensional tracker of the evolving gravitational field quotienting out all 3-dimensional gauge-dependent quantities
\cite{cbbp2004}. In contrast with the eigenvalues of the expansion-normalized extrinsic curvature/shear which is a coordinate dependent measure of the deformation of the time coordinate surface, the eigenvalues of the Weyl tensor are spacetime invariants independent of the properties of the foliation and/or coordinates used to study a given spacetime. The speciality index is a particularly useful combination of those invariants which is automatically ``expansion-normalized" and closely connected to the Petrov classification of the Weyl tensor, an important geometrical tool for classifying spacetime curvature which has found useful application first in numerically generated spacetimes where it is helpful in identifying gravitational radiation.

During a phase of velocity-term-dominated evolution, the Weyl tensor is approximately determined by the extrinsic curvature alone, and hence the scale invariant part of the Weyl tensor is locked to the generalized Kasner indices exactly as in a Kasner spacetime, where the complex speciality index is confined to the interval $[0,1]$ of the real axis. Of course during transitions between velocity-term-dominated evolution where the spatial curvature terms are important, the generalized Kasner indices and the Weyl tensor evolve independently, but the transition between one set of generalized Kasner indices and the next is locked to a transition in the scale invariant part of the Weyl tensor. This idealized mapping, approximated by the 
discrete
BKL map between Kasner triplets, can be reinterpreted as a continuous transition either in the generalized Kasner indices associated with the expansion-normalized shear or in scale invariant part of the eigenvalues of the Weyl tensor. 

For spatially homogeneous vacuum spacetimes, the BKL transition is a consequence of a Bianchi type II phase of the dynamics which can be interpreted as a single bounce with a curvature wall in the Hamiltonian approach to the problem. One can in fact follow this transition in the Weyl tensor directly with an additional first order differential equation which is easily extracted from the Newman-Penrose equations expressed in a frame adapted both to the foliation and the Petrov type of the Weyl tensor.
This type of Weyl transition in the spatially homogeneous Mixmaster dynamics can be followed approximately using the Bianchi type II approximation to a curvature bounce, leading to a temporally isolated pulse pair
in the real and imaginary parts of the speciality index which represents a circuit (``simple loop around the origin") in the complex plane between the two real asymptotic Kasner points on the interval $[0,1]$ of the real axis (a ``complex pulse''), with profiles in time 
qualitatively similar to the pairs of spatial profiles present in expansion-normalized metric/shear conjugate variable pairs near the spikes 
recently found explicitly in spatially inhomogeneous Gowdy cosmologies  by Lim \cite{spike}, 
and numerically or qualitatively by previous authors \cite{ren2001,gar2003,gar2004,gar2007}.
In fact the discontinuity in the asymptotic limit of the speciality index towards the singularity is a gauge invariant characterization of the presence of a 
``permanent true spike" 
in the example discussed in Appendix C.
These 
particular 
spikes seem to describe inhomogeneous discontinuities that seem to arise in asymptotic spatially inhomogeneous BKL dynamics in neighboring curves entering a cosmological singularity.

\begin{figure}[t] 
\typeout{*** EPS figure 1}
\begin{center}
\includegraphics[scale=0.26]{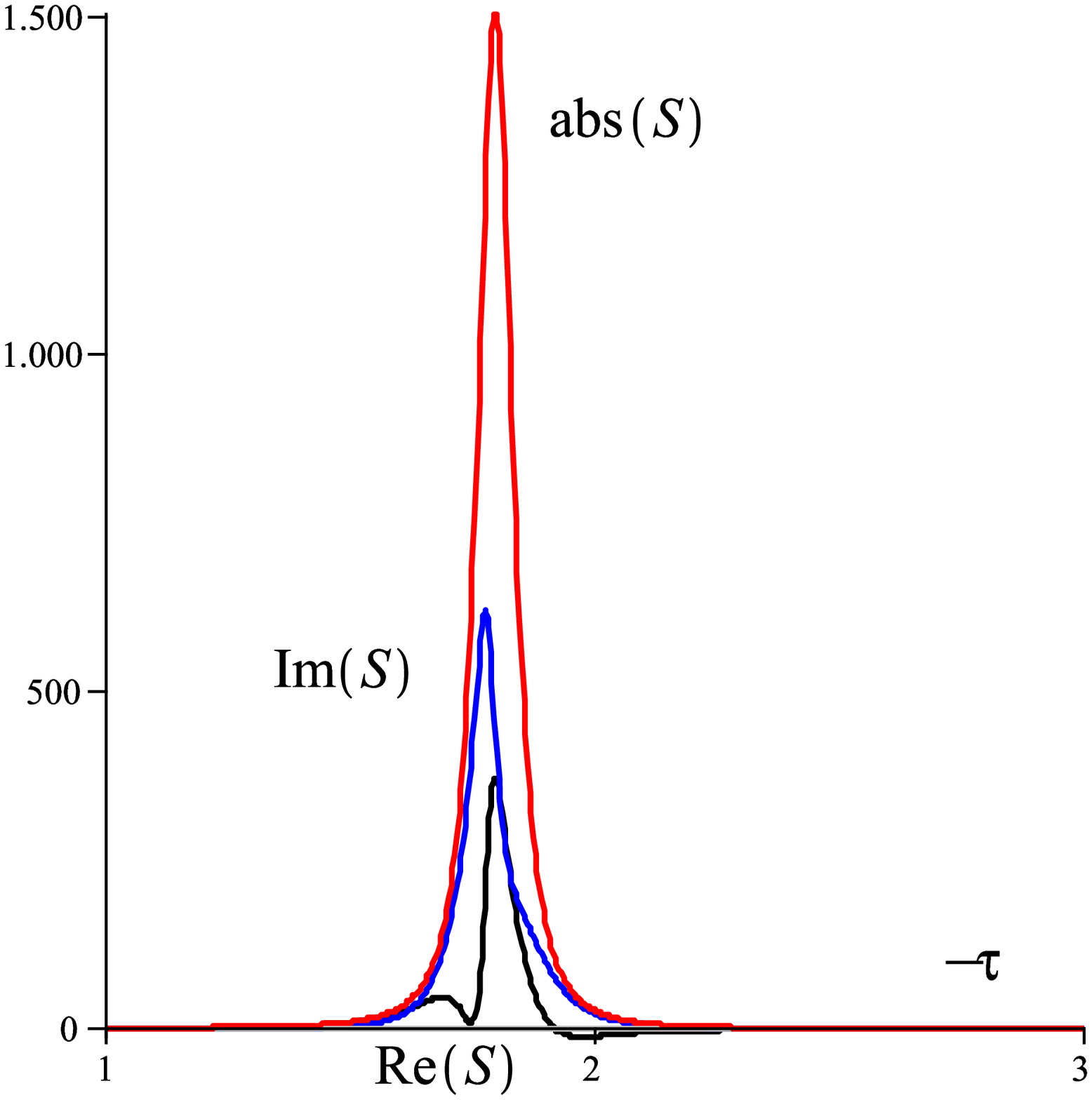} 
\includegraphics[scale=0.22]{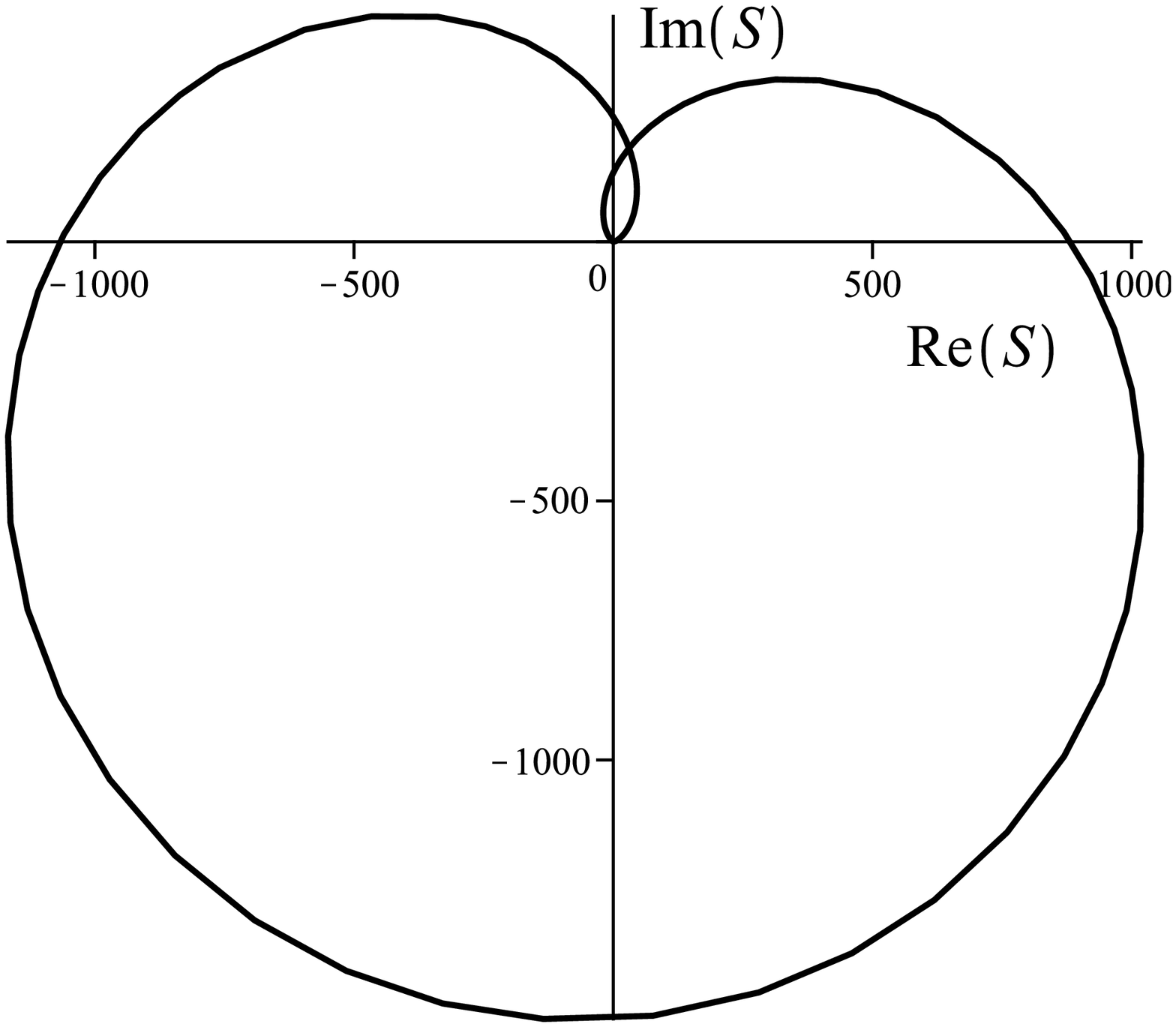} 
\includegraphics[scale=0.15]{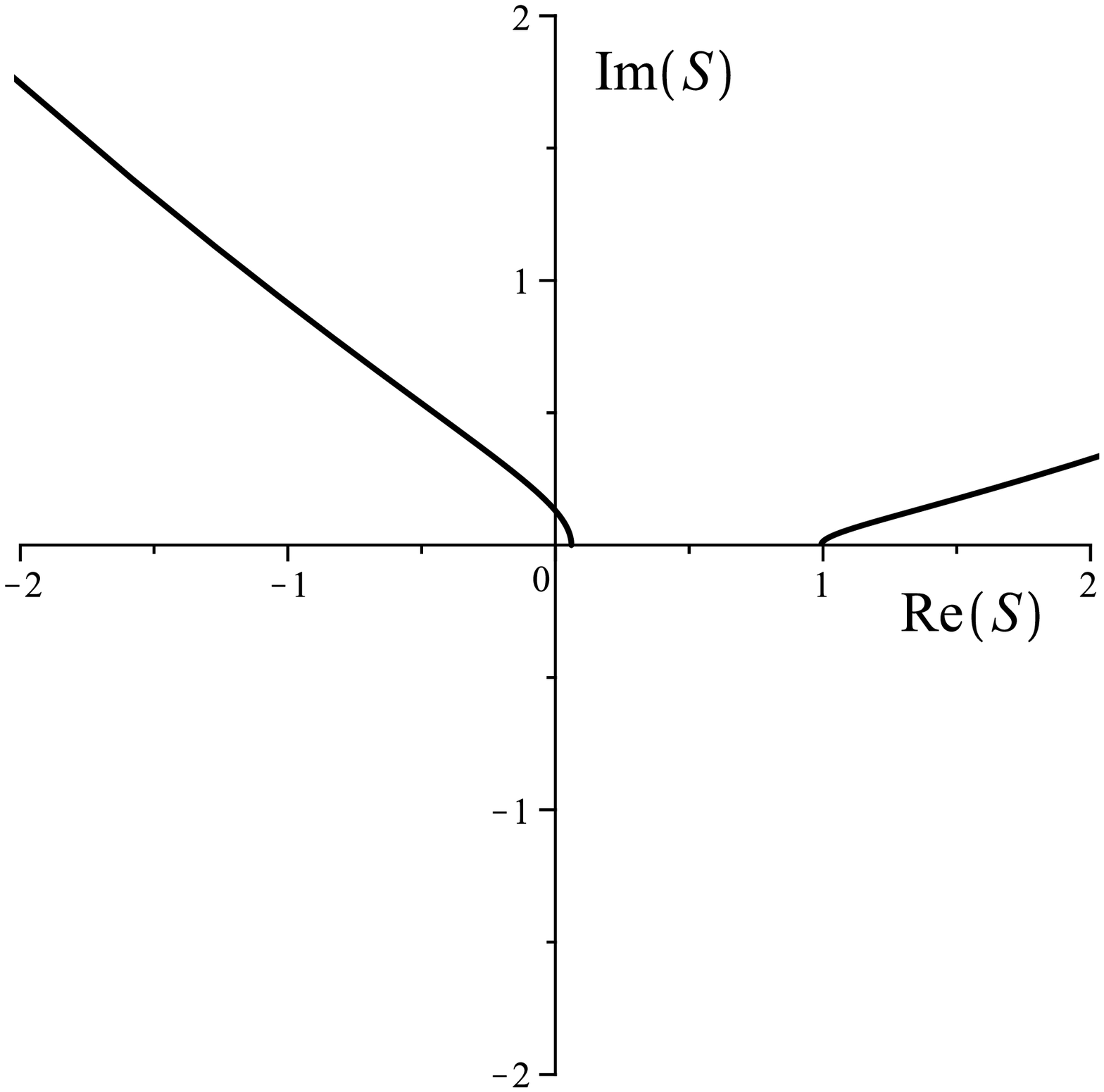}
\\
\hbox to \textwidth{\null\hfill a) \hfill b) \hfill c) \hfill\null}
\end{center}
\caption{
a)
The pulses in the real and imaginary parts of the speciality index for a Bianchi type II Taub solution typical for $u_\infty\to1^+$ where $\mathcal{S}_\infty\to1$, here illustrated with 
$u_\infty=1.1, K=1$  so that $\mathcal{S}_\infty \approx 0.9932$, $\mathcal{S}_{-\infty}=0.0597$. 

b)
The speciality index makes a large cardioid-like circuit in the complex plane starting and ending on the limiting points near 0 and 1, close to the values for normal incidence scattering off the curvature wall which describes the locally rotational type D case.

c)
A closeup of the initial and final points in the complex plane of the previous figure, showing the increasingly horizontal approach to the real axis near the endpoint value 1 on the real Kasner interval.
} 
\label{fig:spikes1}
\end{figure}

\begin{figure}[t] 
\typeout{*** EPS figure 2}
\begin{center}
\includegraphics[scale=0.25]{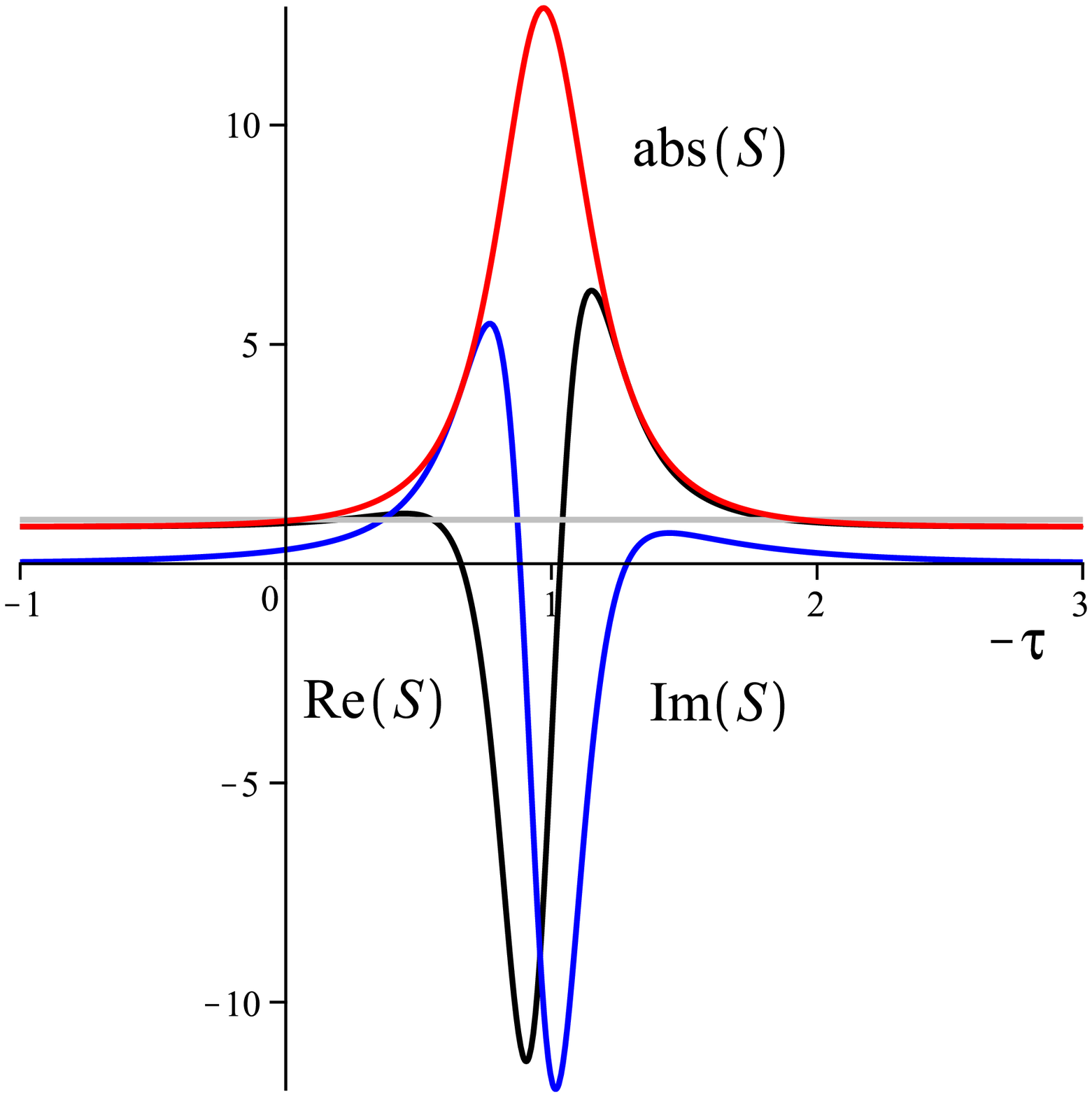}
\includegraphics[scale=0.25]{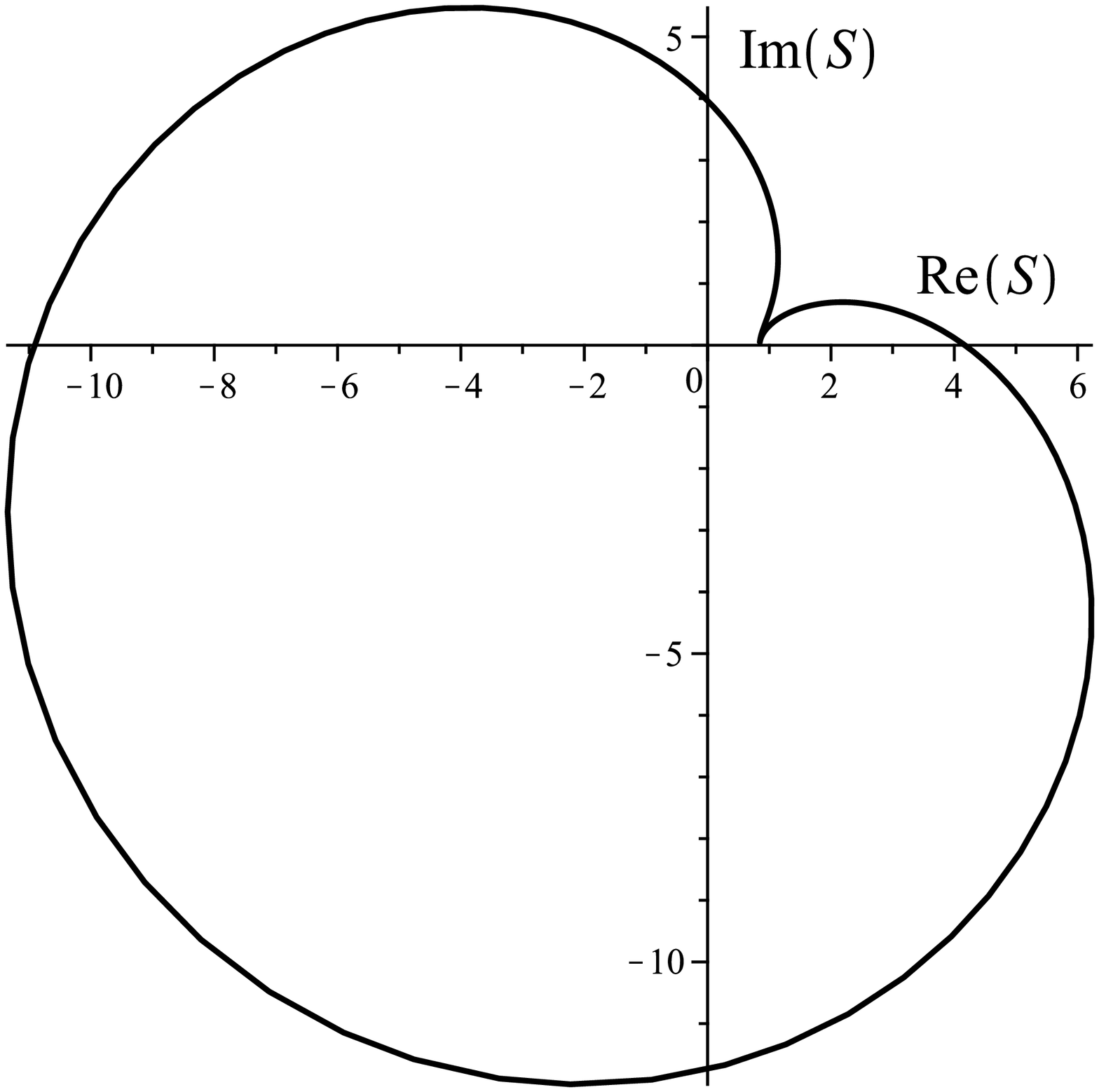}
\end{center}
\caption{
The pulses for a Bianchi type II Taub solution  with $u_\infty=1/(u-1)\approx1.618034, K=1$  so that $\mathcal{S}_\infty = \mathcal{S}_{-\infty} \approx 0.83275$ and the crossing point of the previous cardioid-like circuit now touches the horizontal axis.} 
\label{fig:spikes3}
\end{figure}

\begin{figure}[t] 
\typeout{*** EPS figure 3}
\begin{center}
\includegraphics[scale=0.25]{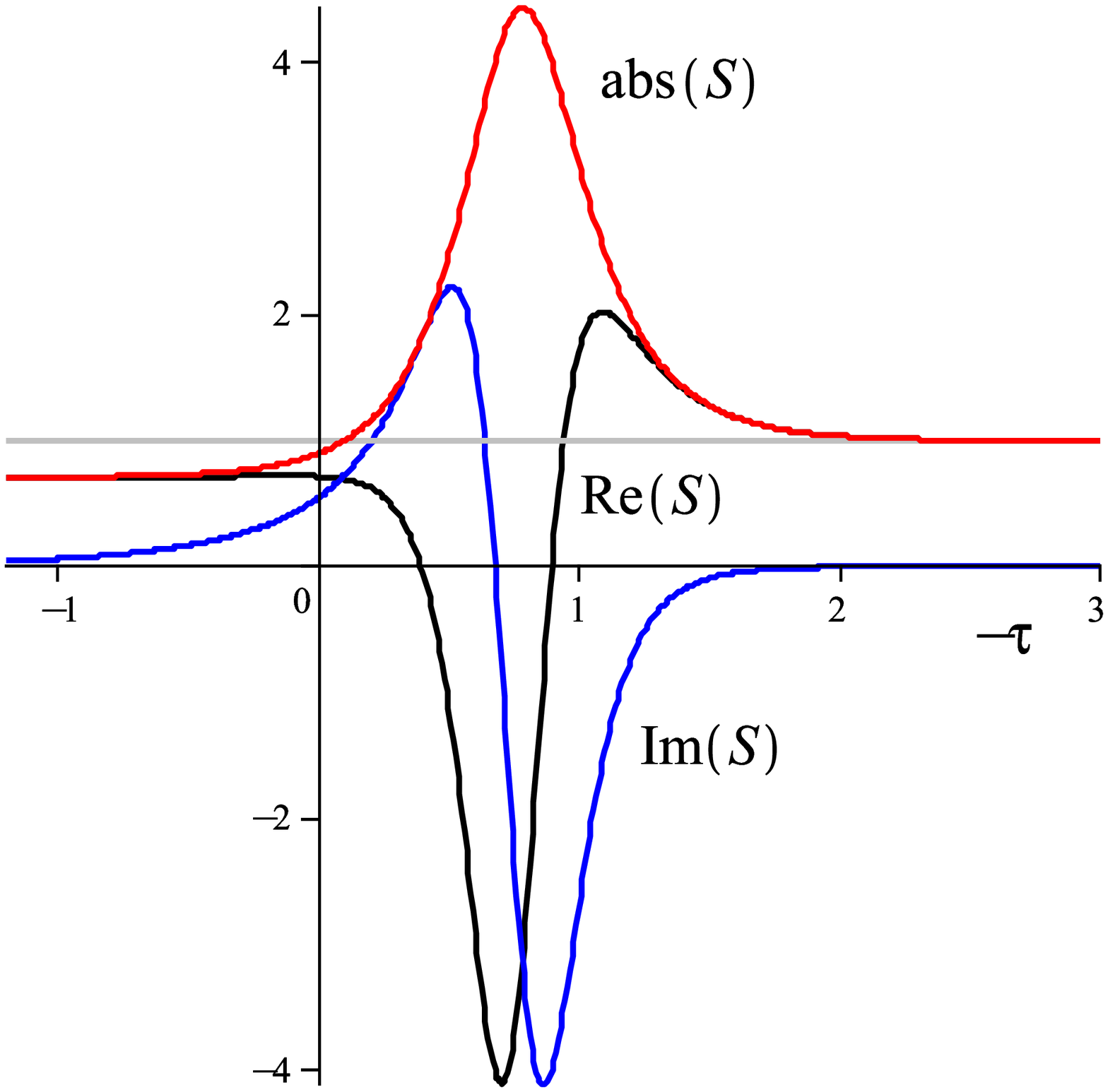}
\includegraphics[scale=0.25]{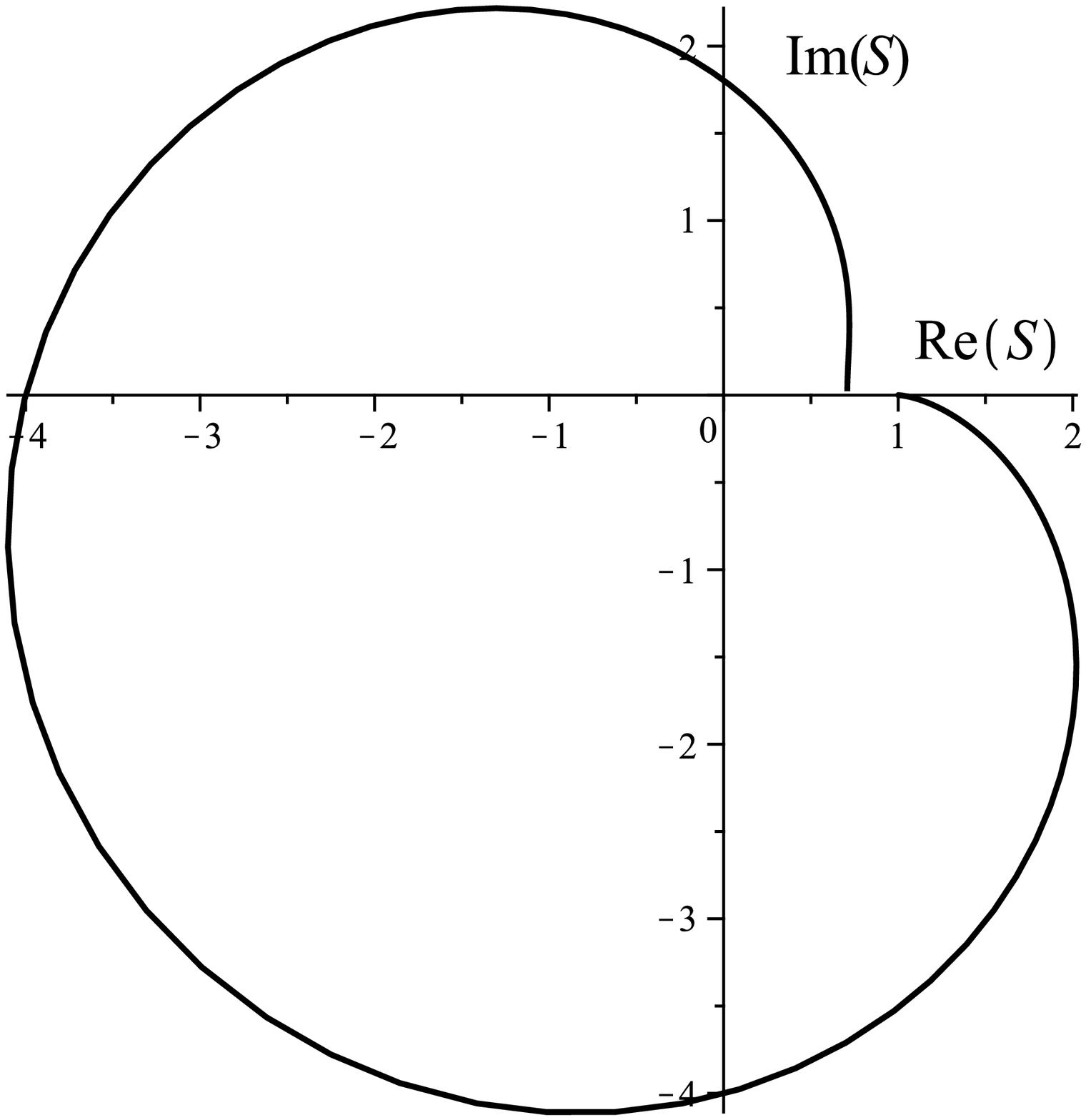}
\end{center}
\caption{
Same as the previous diagram but with $u_\infty=2, K=1$  so that $u_{-\infty}=1 $ and  hence 
$\mathcal{S}_\infty \approx 0.708$, $\mathcal{S}_{-\infty}=1$. Now the right horizontal intercept of the circuit has moved right to be exactly 1 and the tangent line to the circuit there is horizontal.} 
\label{fig:spikes4}
\end{figure}

\begin{figure}[t] 
\typeout{*** EPS figure 4}
\begin{center}
\includegraphics[scale=0.25]{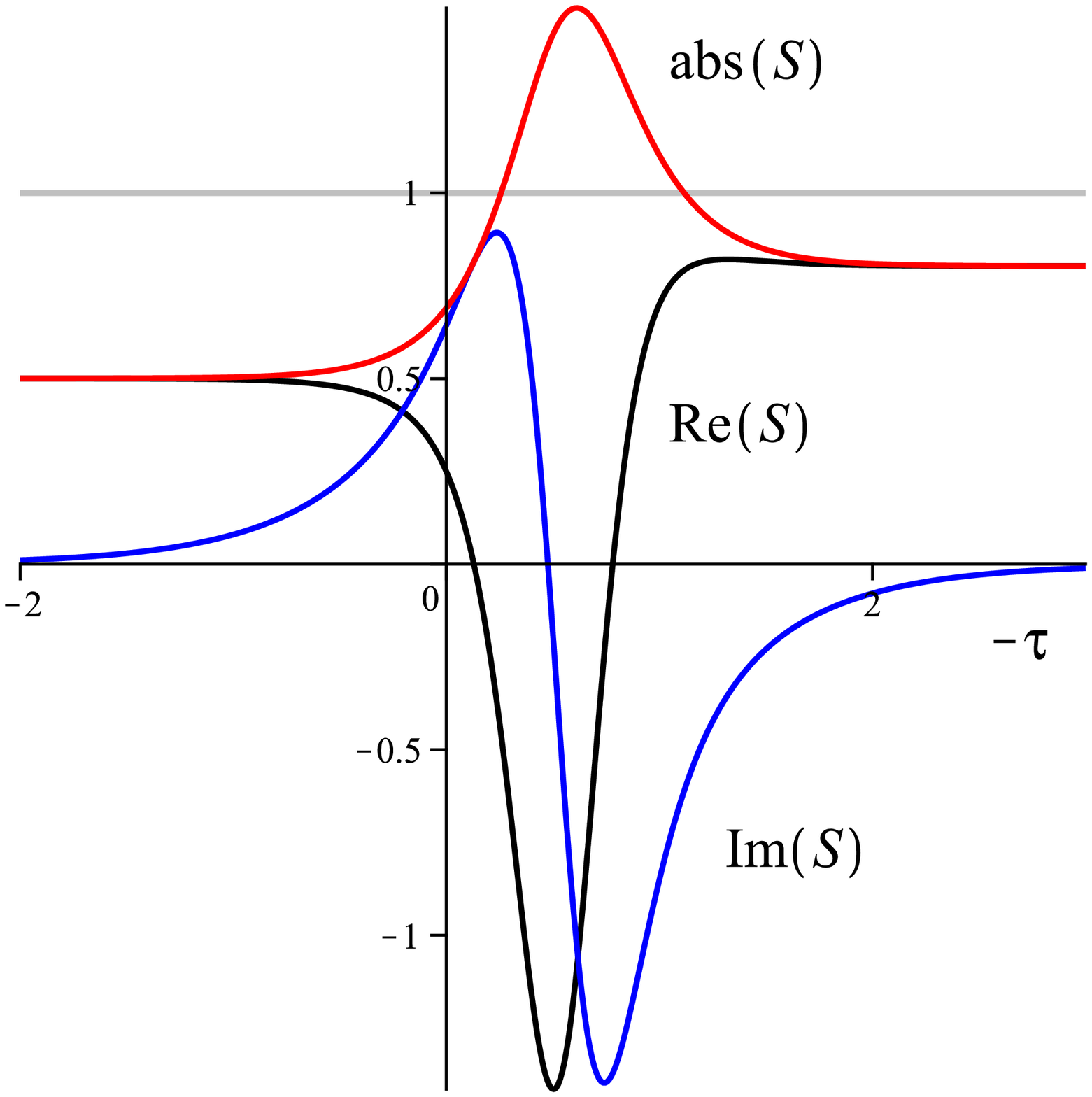}
\includegraphics[scale=0.25]{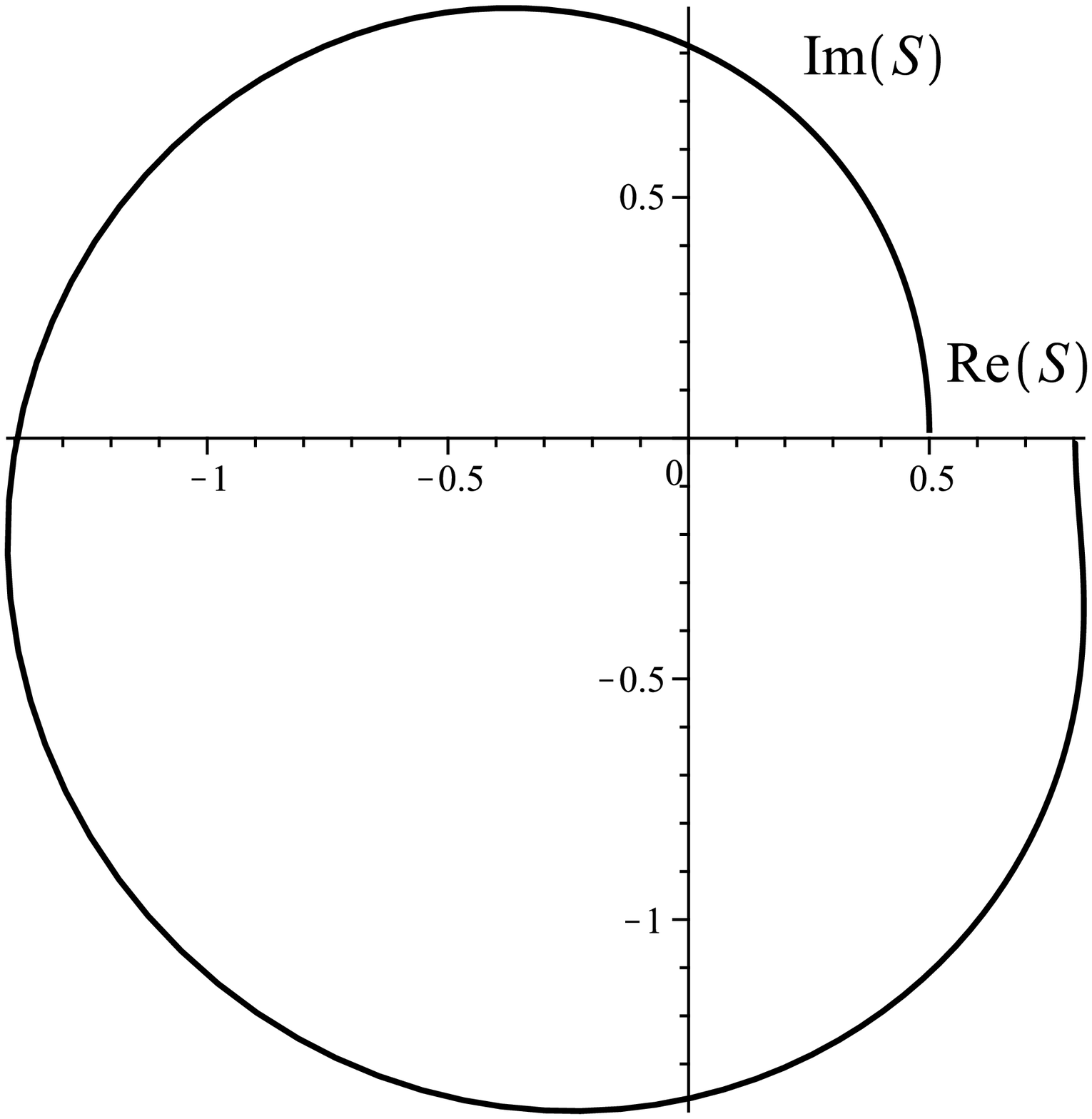}
\end{center}
\caption{
The Bianchi type II pulses typical for $u>2$ illustrated here for  $u_\infty= 2.73$, $K=1$ so that $\mathcal{S}_\infty \approx 0.500$, $\mathcal{S}_{-\infty}\approx 0.8003$.} 
\label{fig:spikes5}
\end{figure}

For the spatially homogeneous Mixmaster universe, the graph of the speciality index versus time serves as a sort of electrocardiogram (EKG) of the  ``heart'' of the Mixmaster dynamics, stripping away all the gauge and frame dependent details of its evolution except for the choice of time parametrization, while the image of the complex curve which produces it is independent of the time as well. This single complex curve in fact represents the plot of the two real scale invariant scalar invariants against each other much like a velocity phase plane plot of velocity versus position during each single oscillation cycle of a mechanical system, here analogous to a single heart beat in this EKG analogy.
Figs.~\ref{fig:spikes1}--\ref{fig:spikes5} illustrate the relationship between the single complex pulse in the speciality index for Bianchi type II vacuum spacetimes and the temporally isolated pulses in its real and imaginary parts graphed versus time. 
Fig.~\ref{fig:bianchi9EKG} illustrates a sequence of three such pulses for a Bianchi type IX vacuum spacetime. The details of both the Bianchi II exact solutions and the Bianchi IX speciality index transitions are left to Appendix B.

The decades of attention given to the Mixmaster dynamics focusing on the extrinsic curvature eigenvalue transitions have really masked transitions in the Weyl tensor. While all of the remarks made below for the spatially homogeneous Mixmaster spacetime extend to the class A Bianchi vacuum spacetimes and, apart from some extra considerations accounting for other sources of spatial derivative terms in the Einstein equations, probably also to inhomogeneous BKL dynamics \cite{inhomBKL,spike,claes2008}, we will limit our detailed discussion to the Mixmaster spacetime alone. However, much of the interest in the Mixmaster spacetime has been its connection to inhomogeneous cosmological dynamics, so it is important to at least demonstrate how the speciality index fits into current investigations of spatially inhomogeneous  exact and numerical cosmological solutions of the Einstein equations, adding a new perspective through which to view this work. In particular 
the speciality index 
reintroduces the spacetime point of view through the 4-dimensional Weyl tensor and its associated scalar invariants in contrast with the coordinate dependent space-plus-time 3-dimensional description primarily used in 
current approaches. 
In some sense this is more in tune with the real spirit of general relativity.

We therefore 
briefly comment on a single example illustrating the spatially inhomogeneous case of $G_2$-symmetric Gowdy spacetimes where exact solutions have been discovered
by Lim \cite{spike} 
with a spike in the gauge-dependent metric variables, leaving most of the details to Appendix C. For these spacetimes, the speciality index is a function only of the two symmetry breaking coordinates $\tau$ (Taub time) and a spatial inhomogeneity coordinate $x$, and 
in the simplest case of permanent spikes
which correspond to a single bounce off a single curvature wall (a ``curvature transition")
can be 
visualized 
as a vertical tubelike surface in 3-space with a vertical strip gap in the real limiting Kasner interval. This can be done
by plotting the speciality index horizontally in the complex plane versus the vertical spatial inhomogeneity coordinate for all times, thus obtaining a ``spatial pulse surface." 
For fixed  $x$ (horizontal cross-sections of this surface), the speciality index makes a circuit around the origin
in the complex plane starting and ending on the real interval $[0,1]$ corresponding to the asymptotic Kasner indices characterizing the velocity-term-dominated limits as $\tau\to\pm\infty$, and its real and imaginary parts trace out a correlated temporally isolated pulse pair
as functions of time, very similar to what occurs in a simple Bianchi type II spacetime 
(the model for a single bounce of a single spatially homogeneous curvature wall), 
a comparison made in Fig.~\ref{fig:spikes6}. 
Instead for fixed Taub time $\tau$, the speciality index winds around the tubelike surface monotonically in $x$ so that its real and imaginary parts instead realize a spatially isolated correlated pulse 
pair in that spatial variable. 
The ``permanent true spikes" 
of Lim's exact solutions \cite{spike} may be seen as a singularity in this single unified spatial pulse surface  describing the gauge-invariant and scale invariant spacetime information characterizing these solutions of the Einstein equations.
Fig.~\ref{fig:gowdyhelix} illustrates this pulse surface. 

The remaining ``transient true spike" solutions involve two successive such curvature transitions, and hence a marriage of two such tubelike speciality surfaces glued together (two pulses).
Fig.~\ref{fig:gowdytransient} shows a typical speciality plot for a fixed value of $x\neq0$ for a transient spike case where the second pulse is just forming.
For 
even
more complicated $G_2$ symmetric spacetimes exhibiting Mixmaster evolution, this speciality surface becomes much more complicated, with an infinite number of such pulses, with spikes arising in this process presumably in a more complicated fashion. 


\section{Metric and curvature}

The Bianchi IX spacetime is a spatially homogeneous spacetime with isometry group $SU(2)$ acting on the compact spatial hypersurfaces of the natural cosmological foliation by the cosmic time $t$ \cite{ryanshepley}. The vacuum solutions of the Einstein equations for this spacetime lead to 
a natural orthonormal frame and dual frame which correspond to the diagonal form of the metric in a symmetry adapted frame
\beq
\label{metric}
\rmd s^2 =\rmd t^2 -[a(t)\, \omega^1]^2-[b(t)\, \omega^2]^2-[c(t)\, \omega^3]^2 \,, 
\eeq
where the three scale functions are functions only of the cosmic time $t$, and describe the time-dependent spatial anisotropy of the space sections. The signature $+$$-$$-$$-$ is used only so that the Newman-Penrose conventions of Chandrasekhar \cite{chandra} can be followed below.
The spatially homogeneous 1-forms $\omega^i$, $i=1,2,3$, 
the natural orthonormal frame $\{e_{\hat\alpha}\}$, $ \alpha=0,1,2,3$ and associated Newman-Penrose frame $\{\ell,n,m\}$,  
and the field equations are reviewed in Appendix A.

In order to explore how the gravitational field evolves near the big bang or big crunch singularities $abc\to 0$
 it is convenient to introduce a new time coordinate $\tau$ called Taub time \cite{taubtime} by setting
\beq\label{taubtime}
\rmd t = abc\, \rmd \tau
\eeq
which pushes these singularities for which $abc\to0$ out to $\tau\to -\infty$ (big bang) or $\tau\to \infty$ (big crunch),
assuming $\tau$ increases towards the future, 
and hence is more useful than the cosmic proper time which compresses the interesting behavior near these limits. Let a dot denote the cosmic time derivative and a prime denote the Taub time derivative, related to each other by $f' = abc\, \dot f$.

The field equations are also more conveniently expressed in logarithmic metric variables $\alpha$, $\beta$ and $\gamma$, defined by
$a=e^{\alpha}$, $b=e^{\beta}$, $c=e^{\gamma}$,
so that
\begin{eqnarray}
\label{fieldeqs3}\fl
&& \alpha'' -\frac12 \left[ (e^{2\beta}-e^{2\gamma})^2-e^{4\alpha}\right]=0
\,,\ 
 \beta'' -\frac12 \left[ (e^{2\gamma}-e^{2\alpha})^2-e^{4\beta}\right]=0
\,, \nonumber \\
&& \gamma'' -\frac12 \left[ (e^{2\alpha}-e^{2\beta})^2-e^{4\gamma}\right]=0
\,,
\end{eqnarray}
with the constraint equation
\begin{eqnarray}
\label{constraint}
\fl
\alpha' \beta' + \beta' \gamma' + \gamma'\alpha'
 +\frac12 (
 e^{2(\alpha+\beta)}
+e^{2(\beta+\gamma)} 
+e^{2(\gamma+\alpha)} 
)
 -\frac14(e^{4\alpha}+e^{4\beta}+e^{4\gamma}) =0
\,.
\end{eqnarray}
These equations should be evolved backwards in Taub time
if $\tau$ increases towards the future, in order
to study the BKL dynamics which aims towards 
an 
initial singularity customarily set at $t=0,\tau\to-\infty$.


The matrix of mixed spatial extrinsic curvature tensor  components with respect to the spatially homogeneous frame
$
(K^i{}_j)  = -\diag(\dot\alpha,\dot\beta,\dot\gamma)
$
is diagonal (like the metric itself)
and one can define the time-dependent generalized Kasner indices as the diagonal values of the ratio of its components with its trace
\beq
  (K^i{}_j)/K^k{}_k = {\rm diag}(p_1,p_2,p_3) 
=\diag(\alpha',\beta',\gamma')/(\alpha'+\beta'+\gamma')
\,,
\eeq
from which it is clear that the generalized Kasner indices represent the scale invariant part of this tensor's eigenvalues and satisfy
$
  p_1+p_2+p_3 = 1
$.
This quotient extrinsic curvature tensor corresponds to the expansion-normalized shear variables in the Ellis-MacCallum-Wainwright
approach to the dynamics of spatially homogeneous cosmologies \cite{ds}, related to the generalized Kasner indices by
\beq\fl\quad
  (\Sigma^i{}_j) =  (K^i{}_j-\delta^i{}_j K^k{}_k/3)/(K^k{}_k/3) 
 = {\rm diag}(3p_1-1,3p_2-1,3p_3-1) \,.
\eeq

The quadratic Kasner index constraint comes from the scale invariant Hamiltonian constraint
\begin{eqnarray} 
&&\frac{K^i{}_j K^j{}_i - (K^k{}_k)^2}{(K^k{}_k)^2} 
 = p_1^2+p_2^2+p_3^2 -1
\\ 
&=& \frac{  ( e^{2(\alpha+\beta)}
           +e^{2(\beta+\gamma)} 
           +e^{2(\gamma+\alpha)}  )
 -\frac12( e^{4\alpha}+e^{4\beta}+e^{4\gamma} )    
}{(\alpha'+\beta'+\gamma')^2}
 \,.\nonumber
\end{eqnarray}
Under conditions where the right hand side is very small, the second Kasner constraint is satisfied:
\beq
p_1^2+p_2^2+p_3^2 =1 \,.
\eeq
In this case one can simplify the the following expressions in the expansion-normalized shear tensor to find
\begin{eqnarray}\label{sigma-kasner}
\fl\quad
 \Sigma^i{}_j  \Sigma^j{}_k  \Sigma^k{}_i
  &=& (3p_1-1)^3 + (3p_2-1)^3 + (3p_3-1)^3
 = 81 p_1 p_2 p_3 +6
 = 3  \det(\Sigma^i{}_j)  
\,,
\nonumber\\
\fl\quad
 \det(\Sigma^i{}_j)  
  &=& (3p_1-1)(3p_2-1)(3p_3-1)
 = 27 p_1 p_2 p_3 +2
\,.
\end{eqnarray}


\section{The Weyl tensor}

Using the unit normal $e_{\hat0}$ to the foliation, one can introduce
the complex (spatial) tensor (see Eqs.~(3.52), (3.53), (3.62) of \cite{ES}, note the misprint $Q_{ab}$ instead of $Q_{ac}$)
\beq
-Q_{\alpha\beta} =  
(C_{\alpha\gamma\beta\delta} +i\, {}^{\sim}C_{\alpha\gamma\beta\delta})
e_{\hat0}^\gamma e_{\hat0}^\delta
= E_{\alpha\beta}+iB_{\alpha\beta}
\,,
\eeq
associated with the self-dual part of the Weyl tensor, 
where the spatial tensors $E_{\alpha\beta}$ and $B_{\alpha\beta}$ respectively denote the electric and magnetic parts of the Weyl tensor with respect to $e_{\hat0}$.
Note that the sign before $i$ in the expression for the dual of the Weyl tensor depends on the convention chosen for the oriented unit volume 4-form
$\eta$, chosen here following \cite{ES}: $\eta_{\hat0\hat1\hat2\hat3}=1$.

The spatial tensor $Q_{\alpha\beta}$ is symmetric and trace-free, i.e., satisfies the conditions
\beq
Q^\alpha{}_\alpha=0\,, \quad  
Q_{\alpha\beta}=Q_{\beta\alpha}\,, \quad 
Q_{\alpha\beta}u^\beta=0
\eeq
and its nonzero components when expressed in any orthonormal frame adapted to $e_{\hat0}$ form a complex symmetric tracefree $3\times 3$ matrix $(Q_{ij})$, $(i,j=1,2,3)$. The mixed form of this spatial tensor represents a complex linear transformation.
The algebraic type of the matrix $\mathbf Q =(Q^i{}_j)$ provides an invariant characterization of the gravitational field at a fixed spacetime point. This Petrov classification based on the eigenvalues $\lambda_1,\lambda_2,\lambda_3$ of $\mathbf Q$, their multiplicities, and the number of linearly independent eigenvectors leads to certain canonical forms for the matrix $\mathbf Q$ called the normal forms which are listed in \cite{ES}. At most two eigenvalues are independent due to the tracefree condition on the matrix
${\rm Tr}\, \mathbf Q =\lambda_1+\lambda_2+\lambda_3=0$.
This classification is insensitive to the scale of the eigenvalues, and so only depends on their scale invariant part.

The most general Petrov type I corresponds to $\mathbf Q$ being diagonalizable, and a canonical or normal orthonormal frame $e_{\hat\alpha}$ for this type is one in which the matrix is diagonal. This is the case for the natural orthonormal frame introduced above for the Bianchi type IX metric (see Appendix A), where the type I normal frame conditions $\psi_1=\psi_3=\psi_0-\psi_4$ lead to the following expression from Table 4.2 of \cite{ES}
\beq
\label{typeI}
\mathbf Q
=\diag(\lambda_1, \lambda_2,  \lambda_3)
=\diag(\psi_2-\psi_0, \psi_2+\psi_0, -2\psi_2)
\,. 
\eeq
The Petrov type D occurs as the subcase $\lambda_1=\lambda_2$ and permutations, conditions which if maintained in time correspond to the Bianchi type IX Taub spacetime.
The scale invariant part of the Weyl tensor can be expressed various ways.
For example, one can choose some ratio of the two independent Weyl scalars like $z=\psi_4/\psi_0$ or one can pick one of six possible permutations of the eigenvalue ratios like
$\mu=\mu_{23}=\lambda_2/\lambda_3$ and get
\beq
\mathbf{Q}
= \lambda_3\,\diag(-(1+\mu),\mu, 1  ) 
=  \psi_0\, \diag(z-1, z+1, -2z)\,,
\eeq
which leads to the relations
\beq
\fl\quad
 \mu= \frac{\lambda_2}{\lambda_3}
    = \frac{\psi_2-\psi_0}{\psi_2+\psi_0}
    = -\frac{1+z^{-1}}{2}\,,
\quad
  z = \frac{\psi_2}{\psi_0}
    = \frac{\lambda_1+\lambda_2}{\lambda_2-\lambda_1}
    = -\frac{1}{1+2\mu}\,.
\eeq
Note that the choice of $z$ rather than some other ratio constructed from the three eigenvalues is also arbitrary, and subject to a similar action of the permutation group like $\mu$ \cite{bcj2007a,bcj2007b}.

\begin{figure}[t] 
\typeout{*** EPS figure 5}
\begin{center}
\leavevmode\kern-0.5in
\hbox{\vbox{
\includegraphics[scale=0.28]{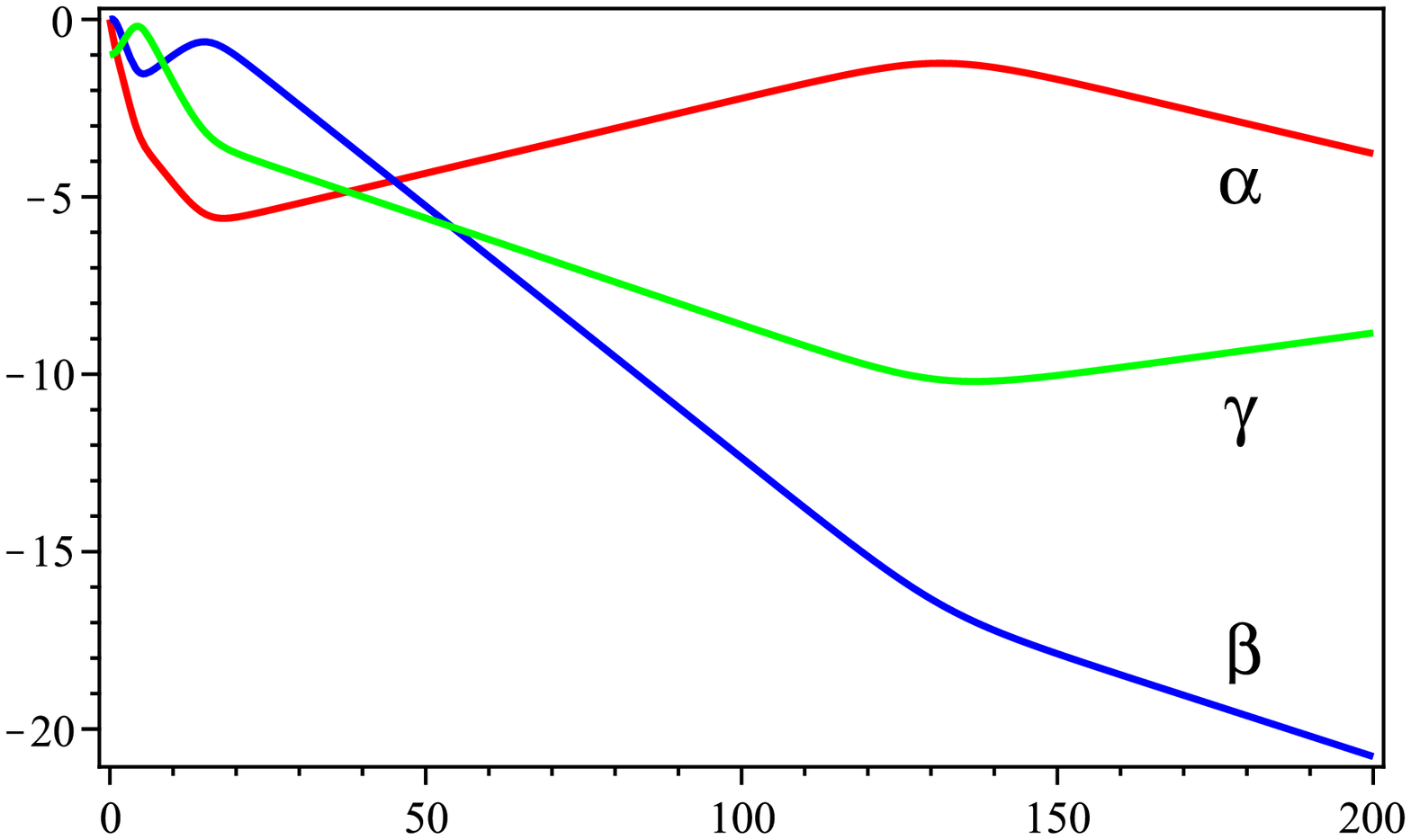}
\\
\includegraphics[scale=0.13]{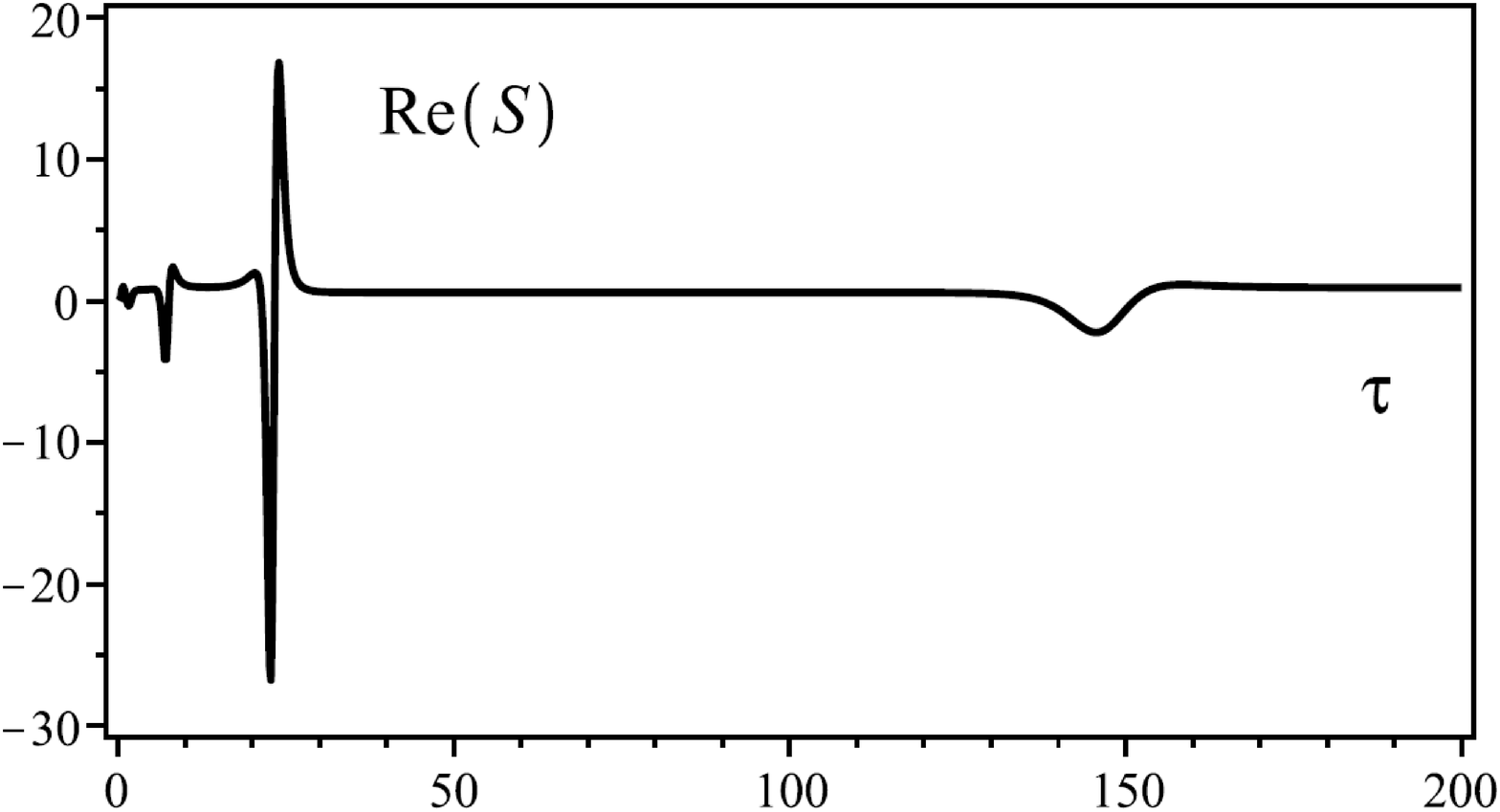}
\\
\includegraphics[scale=0.28]{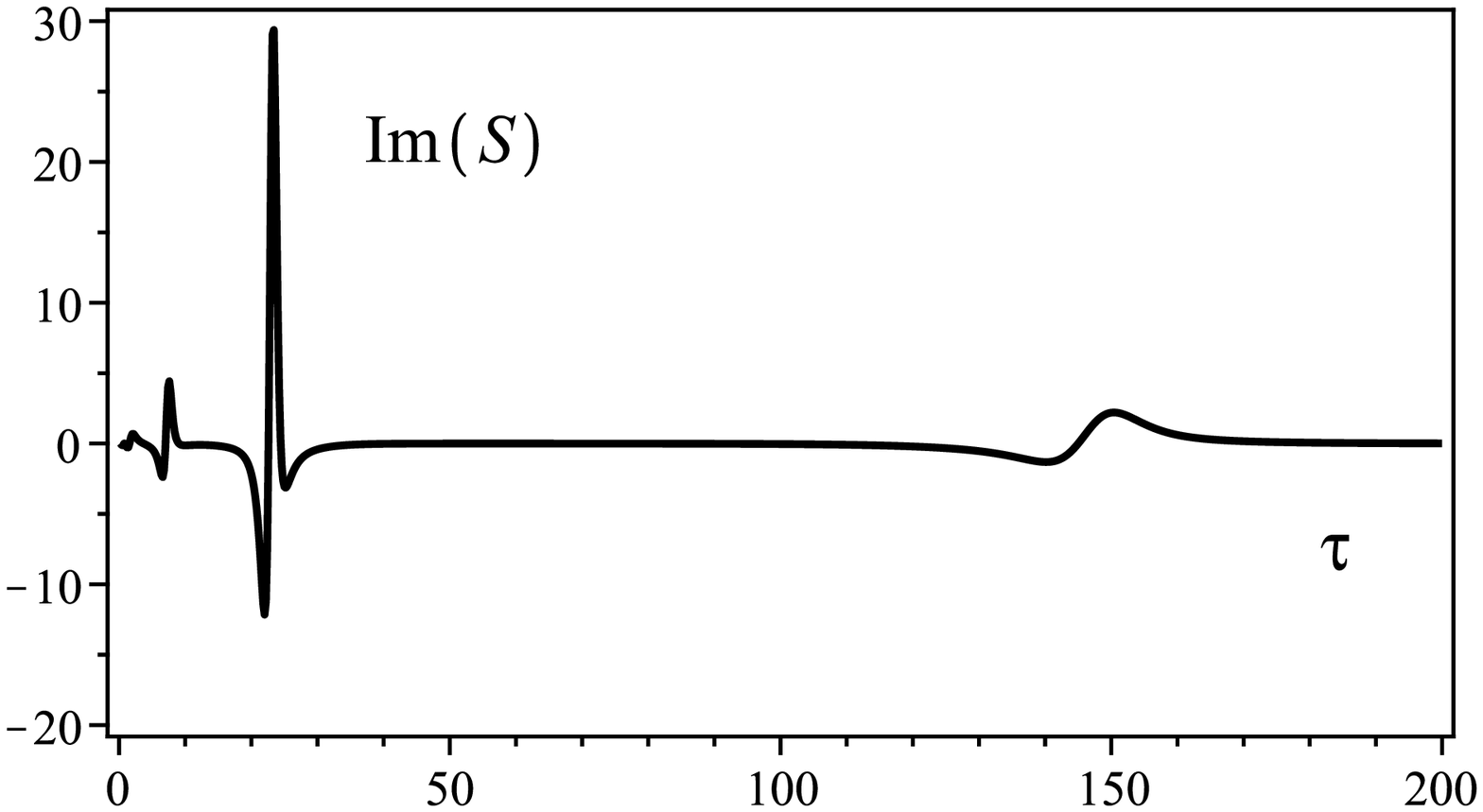}
\\
a)
} 
\kern-3in \vbox{\includegraphics[scale=0.25]{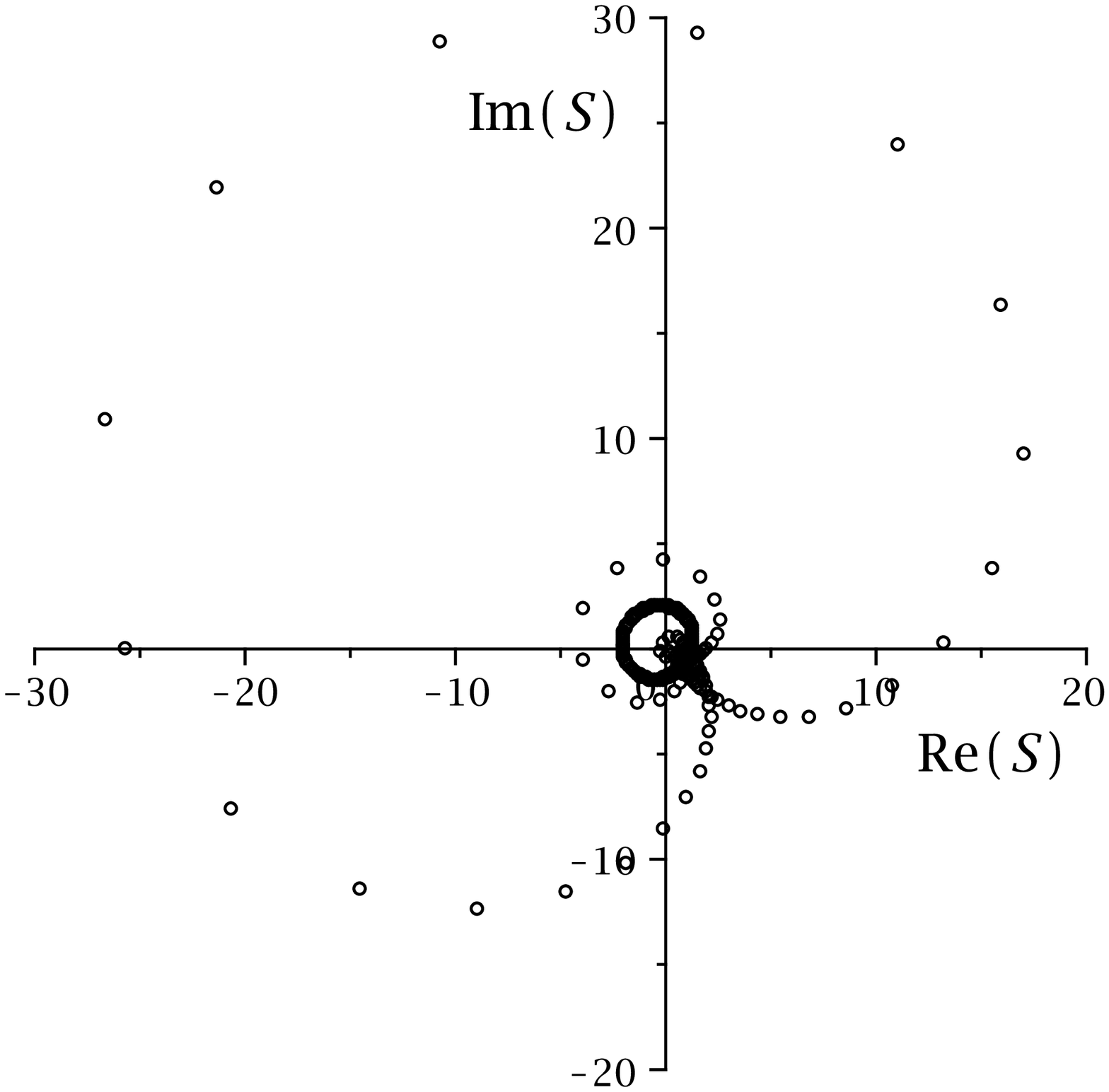}\\
b)}
}
\end{center}
\caption{
a) The pulses in the real and imaginary parts of the speciality index plotted for several consecutive bounces in a Bianchi type IX spacetime. Each collision with a curvature wall generates a pulse in the real and imaginary parts of the speciality index, corresponding to a circuit in the complex plane beginning and ending on the unit Kasner interval of the real axis.

b)
The speciality index plotted as a sampled point plot rather than a smooth curve for these same three consecutive bounces.
} 
\label{fig:bianchi9EKG}
\end{figure}

\section{The speciality index}

The Petrov classification distinguishes between algebraically general spacetimes (type I) and algebraically special ones (types D, II, N, III and O) depending on the degeneracy of eigenvalues and eigenvectors associated with the complex matrix $\mathbf Q$. One way of  determining whether a given spacetime is general or special involves evaluating the scalar invariants determined by the trace of the square and cube of  $\mathbf Q$
\begin{eqnarray}
\fl\quad
\II &=&\frac12 {\rm Tr}\, {\mathbf Q}^2 
=  \frac12 (\lambda_1^2+\lambda_2^2+\lambda_3^2)
= \frac{1}{32}\tilde C^{\alpha\beta}{}_{\gamma\delta} 
  \tilde C^{\gamma\delta}{}_{\alpha\beta}
\,, 
\\
\fl\quad
\JJ &=&\frac16 {\rm Tr}\, {\mathbf Q}^3 
= \frac16 (\lambda_1^3+\lambda_2^3+\lambda_3^3)
= \frac12 \lambda_1 \lambda_2 \lambda_3
= \frac{1}{12\cdot 32}\tilde C^{\alpha\beta}{}_{\gamma\delta} 
  \tilde C^{\gamma\delta}{}_{\mu\nu} 
  \tilde C^{\mu\nu}{}_{\alpha\beta}
\,.\nonumber
\end{eqnarray}

For Petrov types I and then D and II in the top two levels of the Petrov hierarchy, both invariants $\II$ and $\JJ$ are nonzero, while both vanish for types O, N and III in the lowest level of the three level pyramid reprsenting the Penrose specialization diagram (Fig.~4.1 of \cite{ES}).
Any algebraically special spacetime  satisfies
$
\II ^3=27 \JJ ^2 \,,
$
so except for most degenerate types O, N and III in the Petrov hierarchy, one can introduce the so called ``speciality index" \cite{ES} which is a well defined function of the eigenvalues $\lambda_1,\lambda_2,\lambda_3$ by taking the dimensionless ratio (which is therefore independent of the overall scale of the eigenvalues, i.e., scale invariant)
\begin{eqnarray}
\mathcal{S}&=&\frac{27\JJ ^2}{\II ^3}
=6\frac{(\lambda_1^3+\lambda_2^3+\lambda_3^3)^2}{(\lambda_1^2+\lambda_2^2+\lambda_3^2)^3}
= \frac94 \frac{(\lambda_1 \lambda_2 \lambda_3)^2}{(\lambda_1^2+\lambda_2^2+\lambda_3^2)^3}
\end{eqnarray}
having the value 1 for the algebraically special types D and II and which is obviously invariant under any permutation of the eigenvalues. Even though it may not be well-defined for Petrov types  O, N and III, a well-defined limit may exist within a family of special spacetimes, as occurs in the Bianchi type I case of the Kasner family of vacuum spacetimes.

Re-expressing the speciality index in terms of either of the two scale invariant variables $z$ or $\mu$ one finds
\beq
\mathcal{S}
=
\frac{27}{4}\frac{\mu^2(1+\mu)^2}{(1+\mu+\mu^2)^3}
=
\frac{z^2(z^2-1)^2}{(z^2+1/3)^3}
\,.
\eeq
This object is invariant under the permutation group and hence does not depend on the particular ordering of the spatial axes like any of the possible choices for $\mu$ or $z$ do. It is the only independent scale invariant combination of the scalar invariants of the Weyl tensor which is also invariant under a permutation of the eigenvalues.

For Bianchi type I dynamics where the generalized Kasner indices are constant and satisfy the additional quadratic identity, there is only one independent parameter which can be taken to be the scale invariant ratio $u=u_{32}=p_3/p_2$, in terms of which one has the Lifshitz-Khalatnikov Kasner parametrization \cite{LK}
\beq\label{eq:LKK}
p_1=\frac{-u}{1+u+u^2}\,,\
p_2=\frac{1+u}{1+u+u^2}\,, \
p_3=\frac{u(1+u)}{1+u+u^2}\,.
\eeq
One finds that $\mu = u$ is then real \cite{cbbp2004,bcj2007a,bcj2007b}, so that $z$ and $\mathcal{S}$ are also real, and the speciality index has zero imaginary part while the real part is the same expression in $u$ as in $\mu$. 
Indeed the speciality index is then explicitly
\beq\label{Sppp}
   \mathcal{S} = \mathcal{S}_K
= -\frac{27}{4} p_1 p_2 p_3 = \frac{27}{4}\frac{u^2(1+u)^2}{(1+u+u^2)^3} 
= \frac{27(w^2-1)^2}{(w^2+3)^3}
\,,
\eeq
where $w=2u+1$ is the real value of $-z^{-1}=2\mu+1$, a Kasner parameter found useful by Lim \cite{spike}.
Using Eq.~(\ref{sigma-kasner}), this leads to the following relations with the expansion-normalized shear tensor
\beq
 \det( \Sigma^i{}_j) = 2 -4 \mathcal{S}_K\,,
\quad
  \Sigma^i{}_j  \Sigma^j{}_k  \Sigma^k{}_i 
  = 6-12 \mathcal{S}_K \,.
\eeq
One easily sees that in this case the speciality index is confined to the real interval
$0\le \mathcal{S}_K \le 1 $, equaling 1 only when two of the three Kasner indices coincide and are nonzero (type D) and equaling 0 only when two of the three Kasner indices coincide and are 0 (type O). As discussed in \cite{bcj2007b}, the symmetric group of permutations of triplets of numbers acts both on the eigenvalues of the extrinsic curvature and of the Weyl curvature, and hence also on the speciality index, which is invariant under this action. This freedom is useful in adapting the variables $\mu$ or $z$ or $u$ to transitions between velocity-term-dominated phases of the evolution.
Note that the evolution equation (\ref{zdeq}) is easily converted into one for $\mu$ or any of the possible choices for the scale invariant Weyl scalar invariant variable, in terms of which the speciality index can be expressed if this differential equation is integrated numerically in parallel with the usual variables.

\begin{figure}[t] 
\typeout{*** EPS figure 6}
\vglue0cm
\begin{center}
\includegraphics[scale=0.25]{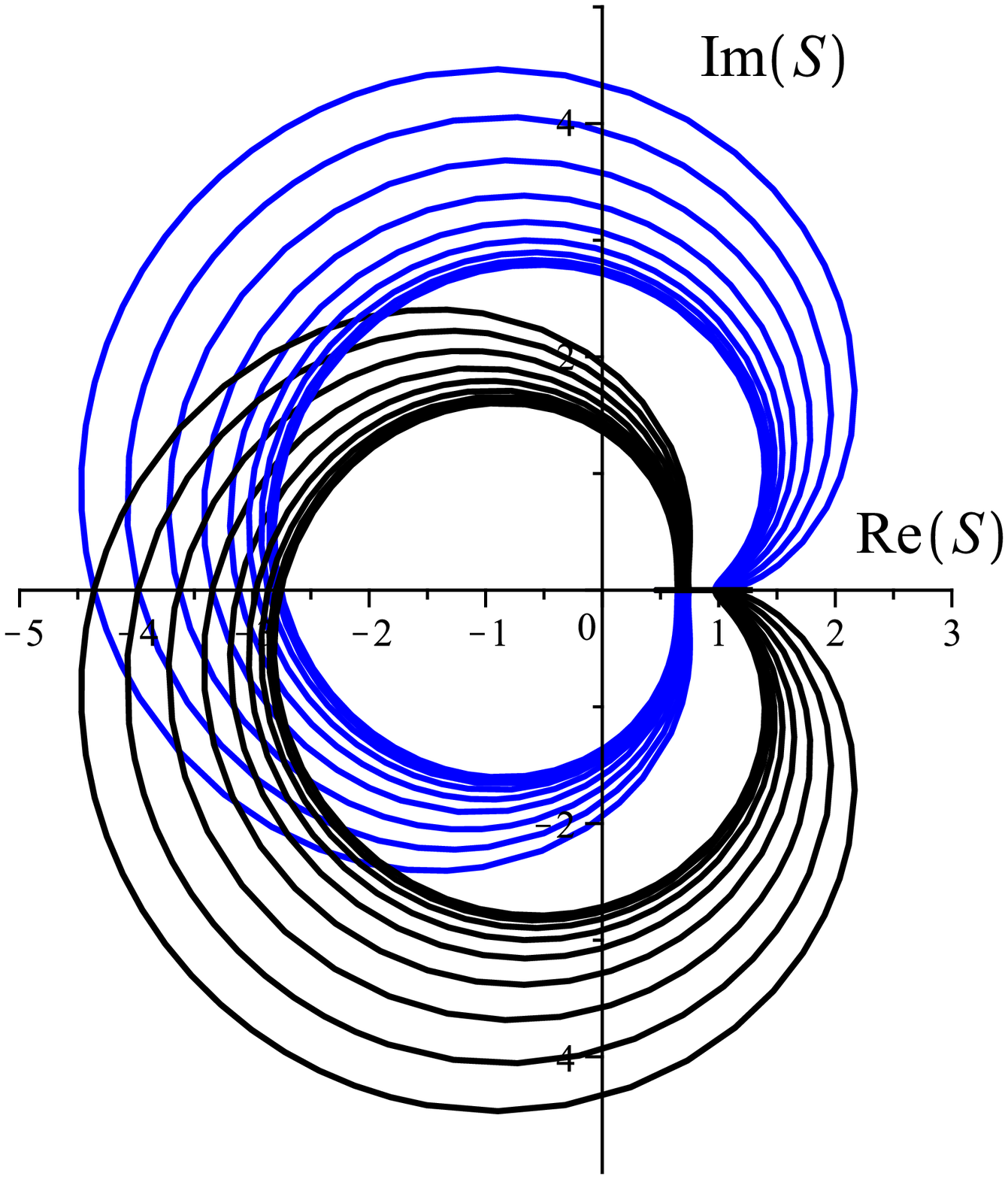}
\includegraphics[scale=0.25]{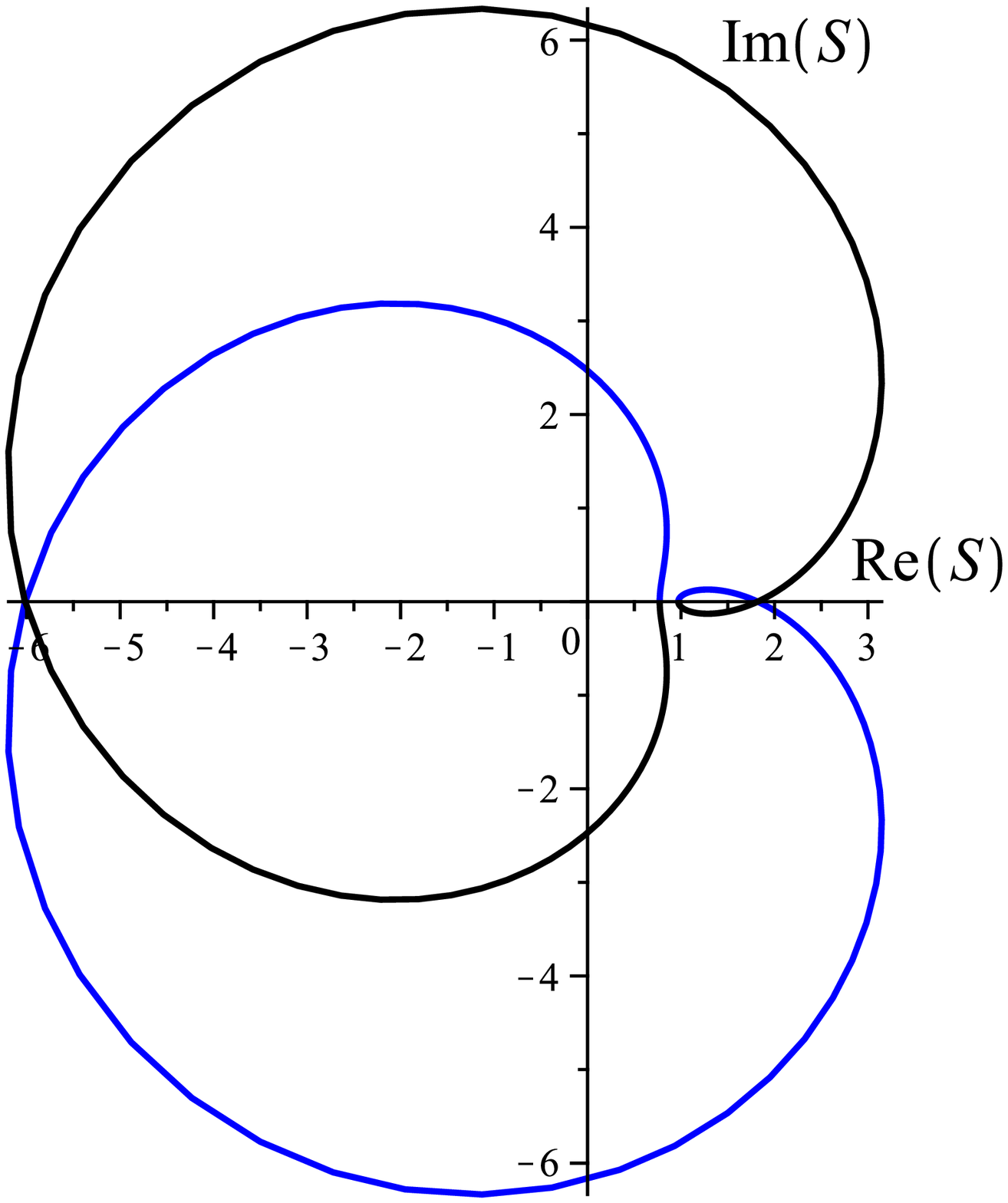}
\end{center}
\null\hfill a) \hfill b) \hfill\null
\caption{
a)
A plot of the speciality index for constant values of $x\neq0$ for one of Lim's Gowdy true spike solutions, shown here for the parameter values
$w=0.1$, $Q_0=1$, $\lambda_2=1$. The cardioid-like circuits lying mostly in the upper half plane correspond to the positive values $x=10^j$ for $j=-5,-4,-3,\ldots,2$ from inside to outside  representing selected spatial points on a logarithmic scale, revealing how the circuit expands with increasing $x>0$. The corresponding curves mostly in the lower half plane are shown for the sign-reversed values of $x$ and arise by a simple reflection across the real axis under the sign change. All circuits for $x\neq0$ start at the common left Kasner point and end at the common right Kasner point, but the single curve for $x=0$ degenerates to line segment from the left Kasner point to an isolated Kasner point further to the left on the unit interval $[0,1]$.

b)
For comparison, the circuit (lower cardioid-like curve, compare with $x<0$ curves to left) and its complex conjugate circuit (upper cardiod-like curve, compare with $x>0$ curves to left) to correspond to time reversal for $x>0$ for
the corresponding Bianchi type II Taub solution seed plot with the same asymptotic Kasner states
$u_\infty=0.55$, $u_{\infty}-1=u=-0.45$
, where $w=2u+1$.
} 
\label{fig:spikes6}
\end{figure}

The speciality index is not only a scale invariant scalar invariant of the Weyl tensor, but during velocity-term-dominated phases of the evolution it also determines the scale of the scalar invariants of the Weyl tensor in a simple way. The simplest combination of the three Weyl eigenvalues which is invariant under the permutation group and is representative of their scale is their product 
$\mathcal{T}=\lambda_1 \lambda_2 \lambda_3$.
For the standard Bianchi type I metric variables where constants have been absorbed into the definition of the spatial coordinates
\begin{eqnarray}
\fl\quad
(a,b,c) = (t^{p_1},t^{p_2},t^{p_3}) = (e^{p_1\tau}, e^{p_2\tau}, e^{p_3\tau})\,,
\quad
 a b c = t^{p_1+p_2+p_3} = t = e^\tau\,,
\end{eqnarray}
this scalar is simple \cite{cbbp2005}
\begin{eqnarray}
\fl\quad
(\lambda_1, \lambda_2, \lambda_3)
&=& 
    t^{-2} p_1 p_2 p_3 \left(\frac{1}{p_1}, \frac{1}{p_2}, \frac{1}{p_3}\right)
 =  t^{-2} \left(-\frac{4}{27} \mathcal{S}\right) \left(\frac{1}{p_1}, \frac{1}{p_2}, \frac{1}{p_3}\right)
\,,
\nonumber\\
\fl\quad
\mathcal{T}
 = \lambda_1 \lambda_2 \lambda_3 
&=& (p_1 p_2 p_3)^2 t^{-6}
  = \left(-\frac{4}{27} \mathcal{S}\right)^2\, t^{-6} \,.
\end{eqnarray}
It scales by a power of the square root of the absolute value of the spatial metric determinant $abc=t$, with a proportionality constant determined by the speciality index itself. 
Note that all of the algebraic scalar invariants of the Weyl tensor are determined by these two complex numbers $\mathcal{S}$ and $\mathcal{T}$, although in the larger context of Kasner transitions in Mixmaster dynamics one must also take into account the contribution from the change in the rescaling of the spatial axes to achieve the Kasner form of the scale factors.

\begin{figure}[t] 
\typeout{*** EPS figure 7}
\vglue1cm
\begin{center}
\includegraphics[scale=0.25]{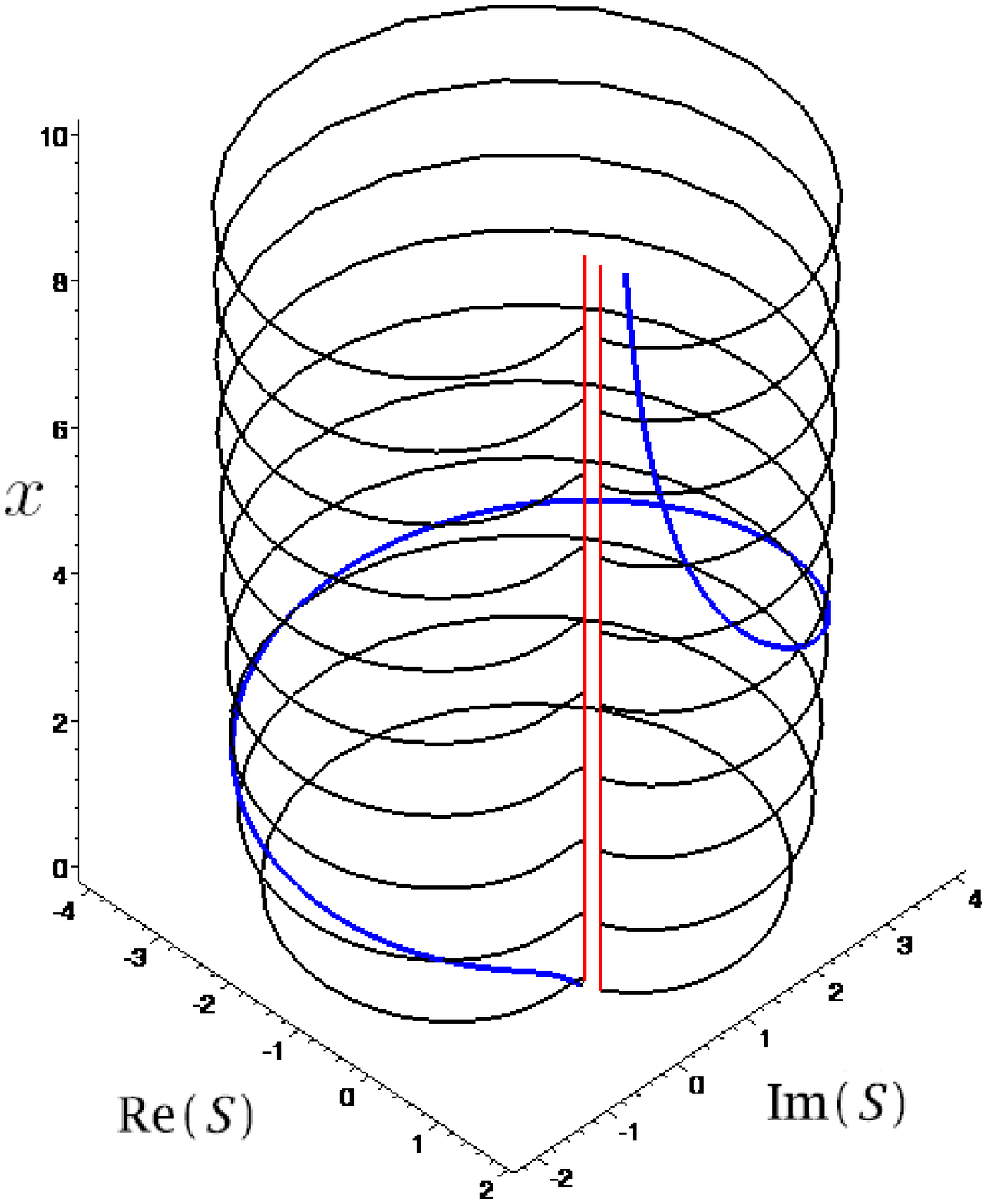} 
\includegraphics[scale=0.25]{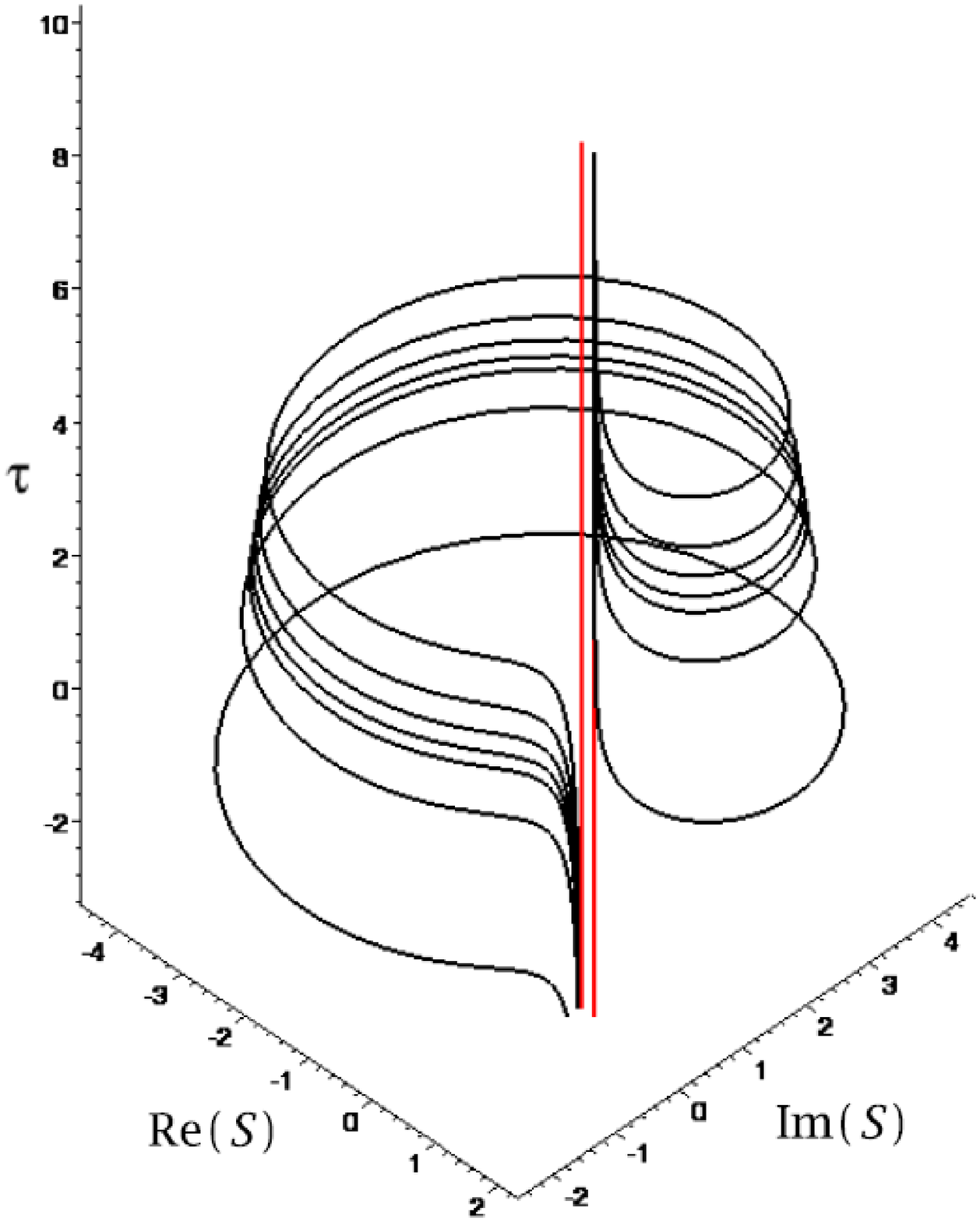}\\
\hfill a) \hfill b) \hfill\hfill\hfill\\
\end{center}
\caption{
Complex plots of the speciality index $\mathcal{S}$ (horizontal) a) versus the spatial coordinate $x>0$ in the vertical direction and b) versus the Taub time $\tau$ in the vertical direction for the Lim's Gowdy true spike solution of the previous figure.

a) Ten horizontal cross-sections $x=1\ldots 10$ are shown, together with the two common asymptotic Kasner values (the two vertical lines through the real axis) which are the beginning and end points for any isotemporal curve, like the typical isotemporal curve  $\tau=3$ shown in the figure, which winds around the tubelike surface with a slowly expanding cross-section as $x$ increases.  
For $x<0$ the symmetry $\mathcal{S}(-x)=\bar{\mathcal{S}}(x)$ reflects the horizontal cross-sections across the real axis, both of which degenerate to a real line segment for $x=0$.

b)
From top to bottom the curves $x=1,2,3,4,5,10,100$ are shown, together with the vertical asymptotic Kasner lines. Each experiences a pulse in a limited time interval which is revealed as spiky behavior in the real and imaginary parts of $\mathcal{S}$.
For $x\le0$ the symmetry $\mathcal{S}(-x)=\bar{\mathcal{S}}(x)$ reflects these curves across the real axis.
} 
\label{fig:gowdyhelix}
\end{figure}

In the Mixmaster context, the Kasner expression $\mathcal{S}_K$ as a function of the generalized Kasner indices or their related Lifshitz-Khalatnikov parameter extends to a time-dependent function on the spacetime locked to the trace of the cube of the expansion-normalized shear tensor. It agrees with the speciality index appromimately only during velocity-term-dominated phases of the evolution.
This shear quantity expressed in terms of the variable $u$ has been used by Garfinkle \cite{gar2007} in the larger context of Gowdy spacetimes.

To summarize, the speciality index is a simple scale invariant scalar invariant of the Weyl curvature tensor which, unlike any of the possible $\mu$ or $z$ variables whose values are shuffled around six different regions of the complex plane by the permutation group \cite{bcj2007b}, is independent of the ordering of the spatial axes and has real values confined to the closed interval $[0,1]$ during velocity-term-dominated phases of the evolution. During a transition between velocity-term-dominated phases of the evolution in the BKL dynamics
induced by a collision with a spatial curvature wall, it exhibits a complex pulse which joins its initial and final real values. In contrast, all of the variables used to describe the evolution itself are functions of the frame-dependent 3-dimensional metric or extrinsic curvature tensor components. In other words, the speciality index is in many senses a privileged spacetime curvature scalar which tracks the evolution, and whose graph (real and imaginary parts) provides a sort of electrocardiogram of the heart of the Mixmaster universe independent of the body of 3-dimensional gauge-dependent variables which house it, figuratively speaking.

Note that we can also establish a correspondence between the speciality index of the exact Bianchi type IX evolution and the approximate one
defined  in terms of the Gauss map of the discrete approximate BKL dynamics through successive values of $u$ \cite{cbbp2004}.
During an velocity-term-dominated phase of the evolution between ``collisions with curvature walls," the real part of the speciality index is approximately constant and the imaginary part is approximately zero,  and the time interval between such collisions corresponds to a single point in the discrete dynamics. 
One can thus directly compare the 4-dimensional information in the numerically evaluated speciality index with the sequence of discrete BKL values labeled by the number of iterations $n$, as well as determine the Taub time interval between the collisions. 
Since both speciality indices are four-dimensional, we can establish a direct correspondence which allows one to easily estimate some of  the relevant parameters (the $p_i$'s and the relation $n\leftrightarrow\Delta\tau$) from $\mathcal{S}$ from the numerical evolution, giving an alternative procedure with respect to the one implemented in the past by Berger \cite{berger1994,berger1997}.

\begin{figure}[t] 
\typeout{*** EPS figure 8}
\vglue1cm
\begin{center}
\includegraphics[scale=0.25]{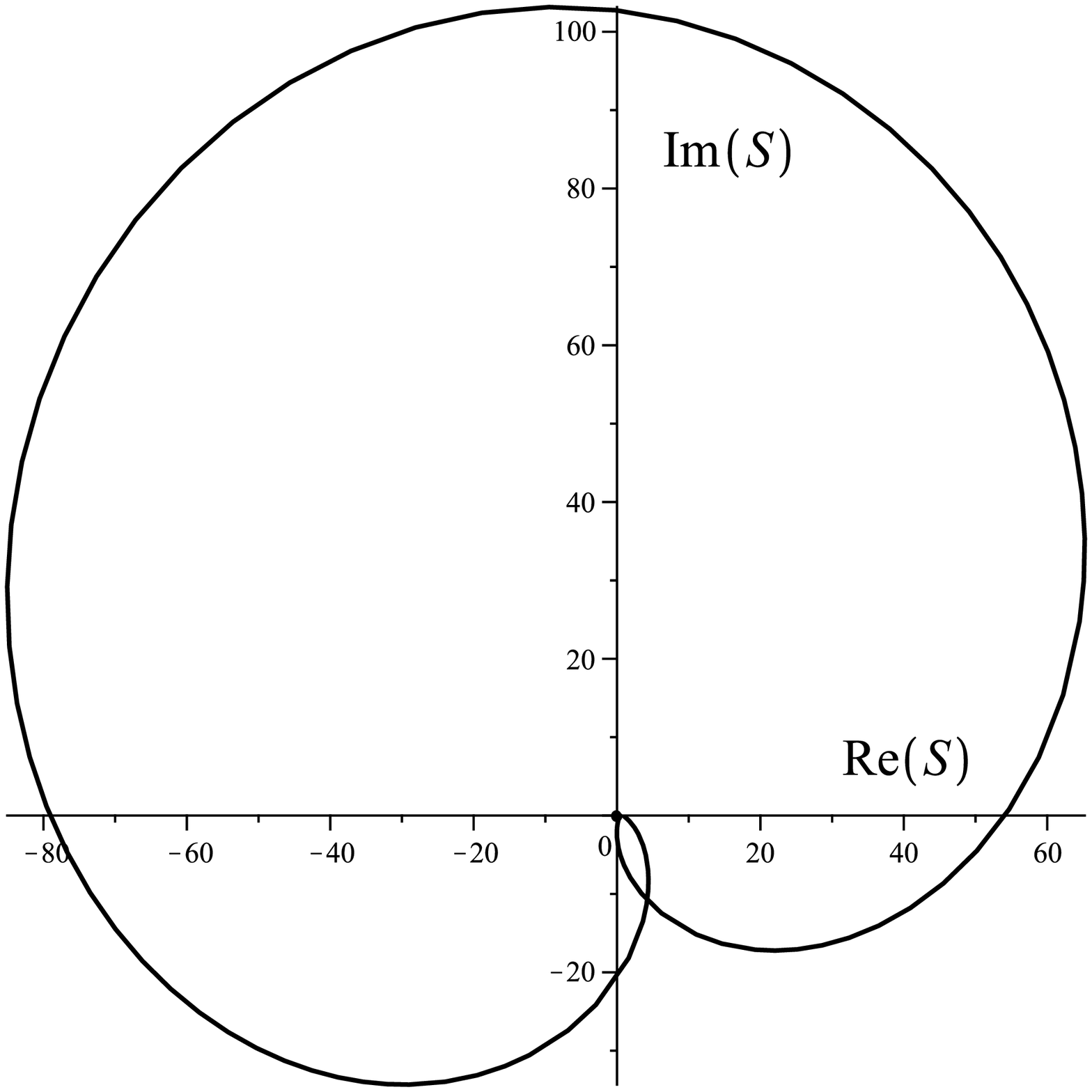} 
\includegraphics[scale=0.25]{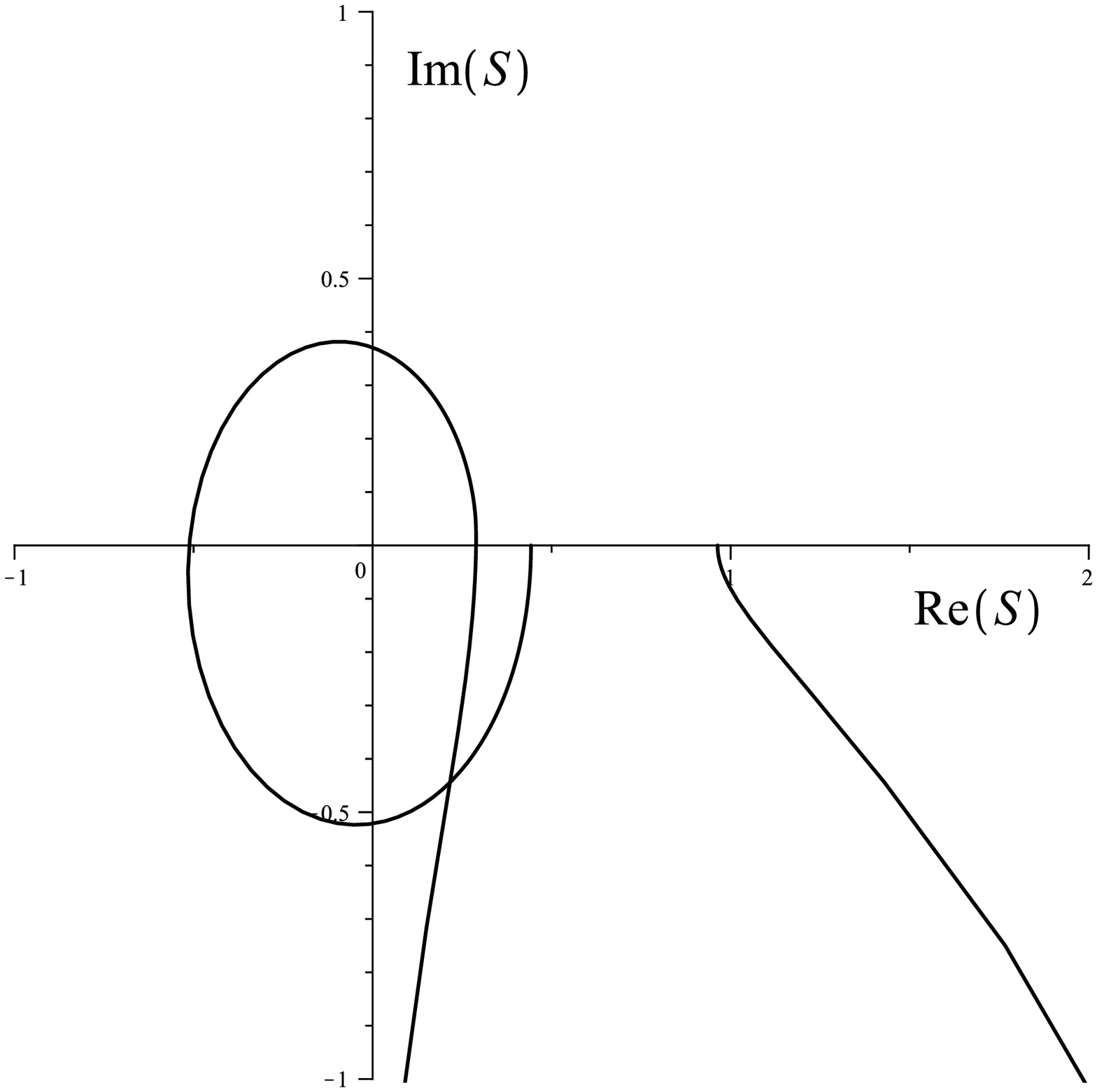}\\
\hfill a) \hfill b) \hfill\hfill\hfill\\
\end{center}
\caption{
A plot of the speciality index for the transient true spike solution of Lim for $w=1.5$ and $x=10$ typical of the case $x\neq0$ away from the location of the spike at $x=0$, with a closeup showing the formation of the second pulse near the origin for $w>1$ which grows relative to the first pulse as $w$ increases.
In the direction towards the singularity, the first pulse begins at the right most Kasner point, making a large clockwise circuit of the origin passing through the left most Kasner point, followed by a tiny counterclockwise circuit of the origin coming to rest at the intermediate Kasner point.
} 
\label{fig:gowdytransient}
\end{figure}

\section{Gravitational waves?}

To some extent, interpreting the Bianchi type IX spacetime as a closed gravitational wave, no matter how convincing the arguments
by Wheeler and Grishchuk, Doroshkevich and Yudin and reviewed by King \cite{king}, who fills in the details omitted by the others, is a debate more about semantics than substance.
The common lore in general relativity is that gravitational radiation is associated with the Weyl curvature tensor, but when no asymptotically flat background is available for comparison (or no short wavelength perturbation of a cosmological model is being considered), it is not really clear how to characterize what behavior in the Weyl tensor can be interpreted as gravitational radiation. All of these discussions rely heavily on special forms for the spatial metric in analogy with the two independent polarizations of gravitational waves on less ambiguous background spacetimes.

However, the speciality index in the context of Mixmaster dynamics seems to be an ideal candidate for an object that is independent of all 3-dimensional quantities except the choice of time in terms of which it is expressed and which can be argued in some sense tracks a gravitational wave evolving in time.
In the asymptotic BKL dynamics, a single collision with a curvature wall generates a complex pulse in this quantity in between intervals of constant real values which change with each collision, as modeled by the exact Bianchi type II vacuum Taub solution discussed in Appendix B. In the generic Mixmaster dynamics far from the initial (or final) singularity where the universe point is simultaneously interacting with all the curvature walls, each one is generating overlapping such pulses which form the profile of the speciality index graph versus time, but in the approach towards the initial singularity, one has a series of temporally isolated pulse transitions between approximately constant behavior as illustrated in Fig.~\ref{fig:bianchi9EKG}, like the pulses in an electrocardiogram.

\section{Concluding remarks}

After an entire book \cite{hobillbook} has been devoted to the spatially homogeneous Mixmaster dynamics including the views of many experts in advanced numerical and qualitative techniques, 
and even more sophisticated analyses have followed in the intervening years,
one might have wondered what more could be said on the subject. 
However, bringing the speciality index into the discussion, originally introduced in the widely different context of modern numerical relativity applied to gravitational radiation problems,
adds an interesting new perspective to a long and familiar story, one which continues even at present as numerical \cite{femlab} and theoretical techniques advance for more general inhomogeneous spacetimes where the speciality index is yet to be exploited \cite{spike,gar2004,claes2008}. 
Although our focus here is primarily on the spatially homogeneous Mixmaster universe,
the exact inhomogeneous Gowdy example found by Lim 
shows that the speciality index gives a gauge-invariant lens through which to view the spike properties of this solution, 
and as such has promise as another important tool for understanding more complicated inhomogeneous numerically-generated spacetimes.

\appendix

\section{Bianchi type IX geometry}

The spatially homogeneous 
orthogonal dual frame 1-forms in terms of which the metric is simply expressed
have the local coordinate form (useful for computer algebra calculations)
$\omega^0 = \rmd t$ and
\begin{equation*}\fl
\omega^1 = \cos \psi \, \rmd \theta +\sin \psi \sin \theta \, \rmd \phi
\,,\
\omega^2 = \sin \psi \, \rmd \theta -\cos \psi \sin \theta \, \rmd \phi
\,,\ 
\omega^3 = \rmd\psi  +\cos \theta \, \rmd \phi 
\end{equation*}
and the orthogonal frame itself as $e_0= \partial_t $ and
\begin{eqnarray*}\fl\quad
\label{ortho}
e_1&=& -\cot \theta \sin \psi \, \partial_\psi
      +\cos \psi \, \partial_\theta
      + \frac{\sin \psi}{\sin \theta}\, \partial_\phi 
\,,\\ \fl\quad
e_2 &=& \cot \theta \cos \psi \, \partial_\psi
      -\sin \psi \, \partial_\theta
      +\frac{\cos \psi}{\sin \theta}\, \partial_\phi  
\,,\
e_3= \partial_\psi\,,
\end{eqnarray*}
where the local coordinates have the ranges $x^0\equiv t\in [0,\infty)$, $x^1\equiv \psi\in[0,4\pi]$, $x^2\equiv \theta\in[0,\pi]$ and $x^3\equiv \phi\in[0,\pi]$.

Rescaling the spatial members of the frame and dual frame yields the natural orthonormal frame $\{e_{\hat\alpha}\}$
and  dual frame $\{\omega^{\hat\alpha}\}$
\beq\fl 
  [e_{\hat 0},e_{\hat 1},e_{\hat 2},e_{\hat 3}]
   =   [\partial_t, a^{-1} e_{1}, b^{-1} e_{2}, c^{-1} e_{3}] 
\,,\quad
[\omega^{\hat 0},\omega^{\hat 1},\omega^{\hat 2},\omega^{\hat 3}]
   =  [dt,a\, \omega^{1}, b\, \omega^{2},c\, \omega^{3}]\,.
\eeq
This in turn has a natural associated complex null transverse \cite{transverse} Newman-Penrose frame \cite{ES}
\beq
\label{NP}
l= \frac{1}{\sqrt{2}}(e_{\hat0}+e_{\hat1})
\,,\
n=\frac{1}{\sqrt{2}}(e_{\hat0}-e_{\hat1})
\,,
\ 
 m = \frac{1}{\sqrt{2}}(e_{\hat2}+i\, e_{\hat3}) 
\,.
\eeq

Letting a dot denote differentiation with respect to $t$,
the vacuum Einstein field equations for this metric in Ricci form reduce to
\begin{eqnarray*}
\label{fieldeqs}
\fl
&& \frac{(\dot a bc)\, \dot{}}{abc} +\frac{1}{2a^2b^2c^2}[a^4-(b^2-c^2)^2] =0\,,\ 
 \frac{(a \dot bc)\, \dot{}}{abc} +\frac{1}{2a^2b^2c^2}[b^4-(c^2-a^2)^2] =0\,, 
\nonumber \\
\fl
&& \frac{(a b\dot c)\, \dot{}}{abc} +\frac{1}{2a^2b^2c^2}[c^4-(a^2-b^2)^2] =0\,,\
 \frac{\ddot a}{a}+\frac{\ddot b}{b}+\frac{\ddot c}{c}=0\,,
\end{eqnarray*}
a consequence of which is the so-called Hamiltonian constraint
\begin{eqnarray}\fl
    &&(\ln a)\,\dot{} (\ln b)\,\dot{} 
 +  (\ln b)\,\dot{} (\ln c)\,\dot{}
 +  (\ln c)\,\dot{} (\ln a)\,\dot{} 
   - \frac{1}{4}(a^4+b^4+c^4) 
   +\frac{1}{2}(a^2b^2 + b^2c^2 + c^2 a^2)
= 0 
\,.\nonumber
\end{eqnarray}

The nonvanishing spin coefficients in the natural Newman-Penrose null frame are given by
\begin{eqnarray}
\fl
\rho&=&-\mu
=-\frac{1}{2\sqrt{2}}\left[  \frac{\dot c}{c} +\frac{\dot b}{b}+i\frac{a}{bc}\right] 
\,,\quad 
\lambda = -\sigma 
= -\frac{1}{2\sqrt{2}}\left[  \frac{\dot c}{c} -\frac{\dot b}{b}+\frac{i}{a}\left(\frac{b}{c}-\frac{c}{b}\right)\right] 
\,,\nonumber \\
\fl
\epsilon&=& -\gamma 
=  \frac{1}{2\sqrt{2}}\left[  \frac{\dot a}{a}+\frac{i}{2}\left(\frac{c}{ab}-\frac{a}{bc}+\frac{b}{ac}\right)\right]  
\,,
\end{eqnarray} 
and the nonvanishing Weyl scalars are
\begin{eqnarray}
\fl\quad
&&\psi_0 =\psi_4
= \frac12 \left[\frac{\dot a \dot b}{ab}-\frac{\dot a \dot c}{ac}
+\frac{a^2-c^2}{a^2b^2} -\frac{a^2-b^2}{a^2c^2}\right]
\nonumber \\
\fl\quad
&&\quad
 +\frac{i}{2abc}\left[\frac{\dot a}{a}(c^2-b^2)+\frac12 \frac{\dot c}{c}(a^2-b^2-3c^2)
%
%
-\frac12 \frac{\dot b}{b}(a^2-3b^2-c^2)
 \right]
\,, \\
\fl
&&\psi_2= 
\frac16 \left[\frac{\dot a \dot c}{ac}
            -2\frac{\dot b \dot c}{bc}
             +\frac{\dot a \dot b}{ab} 
%
%
- \frac{b^2+a^2}{a^2c^2} -\frac{c^2+a^2}{a^2b^2}
              +2\frac{a^4+b^2c^2}{a^2b^2c^2}
        \right]
\nonumber \\
\fl\quad
&&\quad
-\frac{i}{4abc}\left[-2a\dot a + \frac{\dot b}{b}(a^2+b^2-c^2)+  \frac{\dot c}{c}(a^2-b^2+c^2)
 \right]\,.\nonumber
\end{eqnarray}

The Weyl scalars enter the Bianchi identities, which in this case (with $\psi_0=\psi_4$, etc.) imply 
$\delta \psi_0=\delta^*\psi_0=\delta \psi_2=\delta^*\psi_2=0,$
and
\beq\fl\quad 
D\psi_2 =-\lambda \psi_0 +3\rho \psi_2=\Delta\psi_2\nonumber\,, 
\quad 
D\psi_0 =(\rho-4\epsilon) \psi_0 -3\lambda \psi_2=\Delta\psi_0\,.
\eeq
Introducing the scale invariant ratio $z = \psi_2/\psi_0$,
one finds from the previous Bianchi identities
\beq
Dz=\Delta z, \quad Dz=\lambda (3z^2-1)+2(\rho+2\epsilon)z \,.
\eeq
Explicitly the second of these is
\beq\label{zdeq}
\fl\quad
z\,\dot{} 
= -\frac12 \left[\left(\ln \frac{c}{b}\right){}^{\dot{}}-i\frac{c^2-b^2}{abc}\right](3z^2-1)
 -\left[\left(\ln \frac{bc}{a^2}\right){}^{\dot{}}+i\frac{2a^2-c^2-b^2}{abc} \right]z
\eeq
Of course all of these equations are easily re-expressed in terms of the logarithmic metric variables and the Taub time.

Finally the two scalar invariants of the Weyl tensor whose ratio defines the speciality index have the following expressions in terms of the Newman-Penrose curvature quantities
\beq\fl\quad
I = \psi _0\psi _4-4\psi _1\psi_3+3\psi _2^2
\,,\quad 
J = \psi_0\psi _2\psi _4-\psi _1^2\psi _4-\psi _0\psi _3^2+2\psi _1\psi _2\psi 
_3-\psi _2^3 \,. 
\eeq

\section{The Taub Bianchi type II vacuum solution and Bianchi type IX bounces}

The metric of the exact Bianchi type II vacuum solution found by Taub \cite{taub} in the Taub time gauge is given by Eq.~(13.55) and Table 8.2 (p.107)  of \cite{ES}, namely our Eqs.~(\ref{metric}), (\ref{taubtime}) with
\begin{eqnarray}
\fl\quad
(\omega^1,\omega^2,\omega^3) &=& (dx^1-x^3\,dx^2, dx^2,dx^3) \,,
\quad 
\rmd \omega^1 = -(-1)\,\omega^2\wedge\omega^3\,,\ 
\rmd \omega^2 = 0 = \rmd \omega^3\,,
\nonumber\\
\fl\quad
(a,b,c) &=& (X^{-1}, X e^{A\tau}, X e^{B\tau}) \,,
\end{eqnarray}
where
\beq
 X^2 = k^{-1} \cosh k\tau\,,\quad
 k= 2 (AB)^{1/2}\,,
\eeq
and the initial cosmological singularity $abc\to0$ occurs at $\tau\to-\infty$.
By defining $\epsilon = \sgn A = \sgn B$ and
\beq
K = (AB)^{1/2}=k/2\,,\
u_\infty = \epsilon (B/A)^{1/2} = -u_{-\infty}\,,
\eeq
one has
\beq
 A = K/u_\infty\,,\
 B = K u_\infty\,,\
 X = \left(\frac{e^{2K\tau}+e^{-2K\tau}}{4K}\right)^{1/2}\,.
\eeq
Note that simple transformation $u_\infty\to u_\infty^{-1}$ has the effect of interchanging the second and third directions, which is a symmetry of the spatially homogeneous frame and hence of the dynamics. The transformation $u_\infty\to-u_\infty$ is equivalent to the time reversal transformation $\tau\to-\tau$.
The value $u_\infty=1$ corresponds to the locally rotationally symmetric Petrov type D case, where the speciality index is identically 1. The value $u_\infty=0$ is clearly not allowed.

With these new parameters one has the Kasner limits
\begin{eqnarray}\fl\quad
&&  [a,b,c] \stackrel{\tau\to\pm\infty}{\longrightarrow} 
[ (4K)^{1/2} e^{\mp K\tau}, 
\quad 
  (4K)^{-1/2} e^{K(\pm 1 +u_\infty^{-1})\tau},  
  (4K)^{-1/2} e^{K(\pm 1 +u_\infty     )\tau}
]
\end{eqnarray}
from which one can read off the asymptotic Kasner indices from the coefficients of $\tau$ in the exponentials divided by their sum
\begin{eqnarray}\fl
  (p_1,p_2,p_3)_{\pm\infty}
   &=& \frac{K (\mp 1, \pm 1 + u_{\infty}^{-1},\pm 1 + u_{\infty} )}{K(\pm1 +  u_{\infty} ^{-1}+ u_{\infty}) }
= \frac{ (- u_{\pm\infty}, u_{\pm\infty} +1, u_{\pm\infty} (u_{\pm\infty}+1) )}{1 + u_{\pm\infty}+ u_{\pm\infty}^2 }
\,.\nonumber
\end{eqnarray}
Thus comparing with Eq.~(\ref{eq:LKK}),
$u_{\pm\infty}$ are seen to be the asymptotic values of the Lifshitz-Khalatikov parameter $u$ for this solution, which represents a single bounce against a curvature wall $W1$ in the Hamiltonian picture as illustrated in Fig.~6 of \cite{bcj2007b}, under which this parameter simply changes sign. 
The locally rotationally symmetric type D Kasner solution corresponding to $u_\infty=0$ corresponds to asymptotic motion parallel to the curvature wall, which is not possible.

The metric variables themselves are then
\begin{eqnarray}\fl\quad
  [a,b,c] & \stackrel{\tau\to\pm\infty}{\longrightarrow} &
[ (4K)^{1/2} e^{p_1\delta_\pm\tau}, 
  (4K)^{-1/2} e^{p_2\delta_\pm\tau},  
  (4K)^{-1/2} e^{p_3\delta_\pm\tau}
] \,,\nonumber
\\ \fl\quad
  abc\, \rmd\tau & \stackrel{\tau\to\pm\infty}{\longrightarrow} &
(4K)^{-1/2}  e^{\delta_\pm\tau} \rmd\tau
= e^{\delta_\pm\tau-\ln((4K)^{1/2}\delta_\pm)} \rmd (\delta_\pm\tau) 
= e^{\tilde\tau_\pm} \rmd \tilde\tau_\pm
\,,
\end{eqnarray}
where 
\begin{eqnarray}
\delta_\pm&=& \pm K ( 1 + u_{\pm\infty} + u_{\pm\infty}^{-1}) \,,
\qquad 
\tilde\tau_\pm = \delta_\pm\tau- \ln((4K)^{1/2}\delta_\pm) 
\end{eqnarray}
is an affine transformation from the Taub time to the canonical Taub time of the corresponding Kasner limits at infinity.

When $0<A<B$ then $u_\infty\in (1,\infty)$ lies in the interval in which the Kasner indices are ordered: $p_1< p_2< p_3$. This is the initial state for the bounce in the reversed time direction towards the initial singularity. To reorder the Kasner indices after the bounce one then does an interchange of the axes which maps $u_{-\infty}=-u_\infty \to -(1+u_{-\infty}) = u_{\infty}-1$ as described in \cite{bcj2007b}. 
The asymptotic value of the speciality index then undergoes the real shift
\beq\label{eq:Sjump}
  \frac{\mathcal{S}_{-\infty}}{\mathcal{S}_{\infty}}
  = \left(\frac{1-u_\infty}{1+u_\infty}\right)^2 \left(\frac{1+u_\infty+u_\infty^2}{1-u_\infty+u_\infty^2}\right)^3 \,.
\eeq
on the Kasner interval $\mathcal{S}\in[0,1]$,
which is a big jump in the speciality index on this interval when $u_\infty$ is just slightly greater than 1, illustrated in Fig.~\ref{fig:spikes1}. The value $u=1$ corresponds to the locally rotationally symmetric case of normal incidence to the curvature wall. When $u_\infty$ increases to the golden mean value $u=1/(u-1)=(1+\sqrt{5})/2\approx 1.618034$ which is a fixed point of the BKL map \cite{cornishlevin}, the self-intersection of the circuit descends to the horizontal axis as shown in Fig.~\ref{fig:spikes3}, and then disappears for larger $u$ values.  When $u$ reaches the value 2, the right horizontal intercept moves to the right end of the Kasner interval [0,1] as shown in Fig.~\ref{fig:spikes4}.
A typical situation for $u>2$ is shown in Fig.~\ref{fig:spikes5}.
Large values of $u_\infty$ lead to small changes in the asymptotic values of the speciality index according to Eq.~(\ref{eq:Sjump}).

Following the evolution of the speciality index through the bounce one sees that the real part roughly undergoes a single large deviation away from the pair of asymptotic values, while the imaginary part undergoes a double peaked S-shaped departure from the axis, with its midpoint zero roughly correlated with the extreme value of the real part pulse. 
This pair very roughly looks like a single extrema away from constant asymptotic values on either side and its derivative graph.
The particular values of the two independent parameters $(K,u_\infty)$ of the Taub solution compress or stretch these pulse profiles horizontally (in time) or vertically (in value) but leave their correlated shapes roughly qualitatively the same. 
In fact the complex speciality index curve and hence the extreme values of its real and imaginary parts and magnitude are completely determined by the Kasner parameter $u_\infty$, while the curve's parametrization in time is determined by the bounce parameter $K$, which stretches/compresses the time parametrization, thus changing the location in time and width of the pulse profiles in time. 

The explicit expressions for the individual Weyl scalar invariants $I$ and $J$ are manageable but not very enlightening. One finds
\begin{eqnarray}\fl
I &=& \frac{K^9}{u_\infty^2 C^9} e^{-6(u_\infty+u_\infty^{-1})K\tau}
      f_1(u_\infty,C,S)\,,
\quad
J = \frac{K^6}{u_\infty^2 C^6} e^{-4(u_\infty+u_\infty^{-1})K\tau}
       f_2(u_\infty,C,S)\,,
\nonumber\\ \fl
\mathcal{S} &=& u_\infty^2\frac{f_3(u_\infty,C,S)}{f_4(u_\infty,C,S)}\,,
\end{eqnarray}
where $f_i$ are complex polynomials and $C=\cosh(2K\tau)$, $S=\sinh(2K\tau)$. Thus the parameter $K$ only rescales the Taub time parametrization of the speciality index curve in the complex plane, so this curve (i.e, its image) only depends on the Kasner parameter $u_\infty$.
However, the plots of the real and imaginary parts of the speciality index versus $\tau$, although they all share the same shape corresponding to $K=1$, are stretched or compressed horizontally by this rescaling parameter. 
In particular the locations in time of the pulses and their widths in these plots are stretched/compressed by this parameter compared to the value $K=1$.

The Taub Bianchi type II speciality index satisfies the following symmetries
\beq
\mathcal{S}(u_\infty^{-1},\tau) = \mathcal{S}(u_\infty,\tau)\,,\
\mathcal{S}(-u_\infty,\tau) = \bar\mathcal{S}(u_\infty,-\tau)\,.
\eeq
The first is an invariance under reflection across the normal line to the curvature wall, while the second extends to a local symmetry the asymptotic relation 
$ \mathcal{S}(-u_\infty,\infty)=\mathcal{S}(u_\infty,-\infty)$
satisfied by the real limiting Kasner formula under the reflection $u\to-u$ across the curvature wall direction which results from the bounce with the wall. 
Alternatively the combined change $(u_\infty,\tau)\to(-u_\infty,-\tau)$ leaves the metric invariant while changing the orientation used in the duality operation, leading to the complex conjugate of self-dual objects.

In the context of Mixmaster dynamics approaching the initial singularity, the temporal pulses in the speciality index are isolated in time. Fig.~\ref{fig:bianchi9EKG} illustrates this behavior for the Mixmaster universe corresponding exactly to $\tau\in[0,200] $ in Fig.~3(a) of the Hobill review in Hobill et al \cite{hobillbook}, during which three wall collisions take place in the recollapse phase of the evolution towards a big crunch. The initial data for this numerical solution are
$(\alpha(0),\beta(0),\gamma(0))=(0,0,-1)$, $(\alpha'(0),\beta'(0),\gamma'(0))=(-1,0.1,0.3417)$; the numerical solutions were easily obtained using the computer algebra system Maple. Each collision results in a circuit of the speciality index in the complex plane, revealing itself as a pulse of correlated temporally isolated pulse transitions between approximately constant baselines in the real and imaginary parts as functions of the time, separated by Kasner phases of the evolution, leading to a profile in time resembling an electrocardiogram.

\section{Lim's true spike Gowdy solutions}

Exactly this paired pulse-like behavior is seen in the expansion-normalized variables in recent work on Gowdy solutions by Lim \cite{spike} (see section 4.5), except that they occur as a function of the spatial inhomogeneity coordinate rather than of the time. In fact these pulses occur both in time and space as a manifestation of a tubelike complex pulse surface present in the speciality index plotted versus one of the two nontrivial coordinates of the Gowdy spacetime. 
The Gowdy line element is 
\beq\fl\quad
\rmd s^2 = e^{(\lambda-3\tau)/2}\rmd \tau^2 -e^{(\lambda+\tau)/2}\rmd x^2
          -e^{P-\tau}(\rmd y+Q\rmd z)^2-e^{-P-\tau}\rmd z^2\,, 
\eeq
where $\tau$ is a sign-reversed Taub time gauge time coordinate for which the initial cosmological singularity occurs at $\tau\to\infty$.

As an example, consider  Lim's exact Gowdy spike solution of the vacuum Einstein equations of section 4.5 of \cite{spike} (with his Eq.~(34) for $Q$ here corrected by dividing its right hand side by 4), 
in which his alternative Kasner parameter $w=2u+1$ appears
\begin{eqnarray}
P&=& 2\tau +\ln ({\rm sech} (w \tau))-\ln [1+(we^\tau{\rm sech} (w\tau)x)^2]\nonumber \\
&&-\ln (2Q_0)\,,\nonumber \\
Q&=& -Q_0 w[e^{-2\tau}+2(w\tanh (w\tau)-1)x^2]+Q_2/4\,,\nonumber \\
\lambda &= & -4\ln ({\rm sech} (w\tau))+2\ln [1+(we^\tau{\rm sech} (w\tau)x)^2]\nonumber \\
&&-(w^2+4)\tau +\lambda_2\,. 
\end{eqnarray}
With the parameter choice $w=0.1,Q_0=1,\lambda_2=1$ and $Q_2=0$, 
this solution belongs to the class $0<w<1$ of ``permanent true spikes," which exhibit a single curvature transition and hence are most similar to the spatially homogeneous Taub solutions of Appendix B.
The initial ($\tau\to-\infty$) and final ($\tau\to\infty$) asymptotic Kasner states in the approach towards the initial singularity common to  all values of the spatial coordinate $x\neq0$ in the single BLK curvature transition correspond here to the parameter values $(w+2,w)=(2.1,0.1)$, for which the speciality index has the corresponding asymptotic values given by Eq.~(\ref{Sppp}) to be 
$(\mathcal{S}_{-\infty},\mathcal{S}_{\infty})$ $\approx$ (0.7716,0.9704).
Instead for $x=0$, the initial and final Kasner states correspond to $(w+2,2-w)=(2.1,1.9)$ and 
the speciality index makes a real hyperbolic tangent-like transition 
in time
from $\mathcal{S}_{-\infty} \approx0.7716$ to 
$ \mathcal{S}_\infty \approx0.6369$, 
leading to a straight line segment along the real axis in the complex plane connecting these two Kasner points.
All the constant $\tau$ curves in the singular speciality index surface pass through this line segment connecting the two components for which $x>0$ and $x<0$.
This discontinuity in the speciality index surface at $x=0$ caused by the discontinuous limit of $\mathcal{S}$ as $\tau\to\infty$ towards the initial singularity is the ``true permanent spike" lending its name to this class of solutions of the Einstein equations, 
and its location is characterized by the vanishing of the imaginary part of the speciality index for all time. 
Fig.~\ref{fig:spikes6} illustrates this situation.

One can plot the complex speciality index horizontally versus either the vertical $x$ or $\tau$ coordinate to obtain a tubelike surface with a gap between the two vertical lines representing the common asymptotic Kasner values of the initial and final states after the inhomogeneous curvature wall bounce.
Fig.~\ref{fig:gowdyhelix}(a) shows a plot of the one bounce speciality circuit in the complex plane (horizontal cross-sections) as a function of the spatial inhomogeneity coordinate $x$ along the vertical axis. Each circuit of the speciality index from the common asymptotic initial and final Kasner values (the two vertical lines through the real axis) corresponds to a temporally isolated
correlated pulse pair
in the graph of the real and imaginary parts versus time, in contrast with the pulses revealed in the expansion-normalized variable pair $(N_-,\Sigma_\times)$ graphed versus $x$ (see below). 
Fig.~\ref{fig:gowdyhelix}(b) shows a plot of the speciality index in the complex plane (horizontal cross-sections) instead as a function of the Taub time coordinate $\tau$ along the vertical axis for a sequence of constant values of $x$. Each curve represents a constant value of the time coordinate, which undergoes a complex pulse during a short time period which changes with the value of $x$. In each case the real and imaginary parts of the tubelike surface plot reveal the pulse behavior, versus the time or spatial coordinate.

Note that plotting $X$ versus $Y$ for a pair of real functions of a single variable is equivalent to plotting their complex combination $X+iY$ in the complex plane. This combination
$N_- +i\Sigma_\times$ of spatial connection and extrinsic curvature in the expansion-normalized metric variables 
(defined by Eq.~(13) of \cite{spike}) 
is reminiscent of the Ashtekar variables \cite{ashtekar1991}, and their individual profiles in space shown 
at the bottom of 
Lim's Fig.~7 \cite{spike} or even in Figs.~1--4 of Rendall and Weaver \cite{ren2001}
reveal the same isolated pulse pair behavior as the temporal profiles of the real and imaginary parts of the speciality index, thus revealing a circuit in the complex combination $N_- +i\Sigma_\times$.
Similar circuits occur in the general discussion of Uggla et al \cite{claes2008}, in plots of some of the 
(expansion-normalized) 
real quadratic Weyl scalar invariants against each other shown in their Fig.~9. 
However, the speciality index plot has the advantage of forcing the Kasner phases of the evolution to sit on the compact interval $[0,1]$ of the horizontal axis.

The remaining Lim ``transient true spike" solutions $w>1$ exhibit two such curvature transitions, and hence the speciality surface has two pulse components, distinguished from the permanent true spikes by having common initial and final Kasner limits for all $x$. 
For both permanent and transient spike solutions, the speciality index is completely insensitive to the so called ``frame transitions" exhibited in the expansion-normalized variables, which are a manifestation of the choice of spatial coordinates. Fig.~8 shows a typical speciality plot for constant $x$ for the case $w=1.5$ which reveals the small second pulse which forms as $w$ increases past the value  $w=1$ and grows relative to the first pulse as $w$ continues to increase. 
The speciality index during the initial ($\tau\to-\infty$), intermediate and 
final ($\tau\to\infty$) Kasner phases corresponding to the values $(w+2,w,2-w)=(3.5,1.5,0.5)$ has the values $(\mathcal{S}_{-\infty},\mathcal{S}_{\rm int},\mathcal{S}_{\infty})$ $\approx$ 
$(0.9635,0.2915,0.4424)$ calculated from Eq.~(\ref{Sppp}).
Details will be studied elsewhere.


\end{document}